\documentclass[12pt,preprint]{aastex}
\shorttitle{LMC}
\shortauthors{Carrera et al.}

\begin{document}


\title{The Chemical Enrichment History of the Large Magellanic Cloud}

\author{R. Carrera\altaffilmark{1} and C. Gallart}
\affil{Instituto de Astrof\'{\i}sica de Canarias, Spain}
\email{rcarrera@iac.es}
\email{carme@iac.es}

\author{Eduardo Hardy\altaffilmark{1}}
\affil{National Radio Astronomy Observatory \altaffilmark{2}, Chile}
\affil{Departamento de Astronom\'{\i}a, Universidad de Chile\altaffilmark{3}, Chile}
\email{ehardy@nrao.edu}

\author{A. Aparicio}
\affil{Instituto de Astrof\'{\i}sica de Canarias, Spain}
\affil{Departamento de Astrof\'{\i}sica, Universidad de La Laguna, Spain}
\email{antapaj@iac.es}

\and

\author{R. Zinn}
\affil{Department of Astronomy, Yale University, USA}

\altaffiltext{1}{Visiting Astronomer, Cerro Tololo Inter-American Observatory.
CTIO is operated by AURA, Inc.\ under contract to the National Science
Foundation.}
\altaffiltext{2}{The National Radio Astronomy Observatory is a facility of the National Science 
Foundation operated under cooperative agreement by Associated Universities, Inc.} 
\altaffiltext{3}{Adjoint Professor} 

\begin{abstract}
Ca II triplet spectroscopy has been used to derive stellar metallicities for individual stars in four LMC fields
situated at galactocentric distances of 3\arcdeg, 5\arcdeg, 6\arcdeg\@ and 8\arcdeg\@ to the north of the Bar. Observed metallicity
distributions show a well defined peak, with a tail toward 
low metallicities. The mean
metallicity remains constant until 6\arcdeg\@ ([Fe/H]$\sim$-0.5 dex), while for the outermost field, at
8\arcdeg, the mean metallicity is substantially lower than in the rest of the disk ([Fe/H]$\sim$-0.8 dex). The combination
of spectroscopy with deep CCD photometry has allowed us to break the RGB age--metallicity degeneracy and
compute the ages for the objects observed spectroscopically. The obtained age--metallicity relationships for our four fields are statistically indistinguishable. We conclude that the lower mean metallicity in the outermost field is a consequence of it having a lower fraction of intermediate-age stars, which are more metal-rich than the older stars. The disk age--metallicity relationship is similar to that for clusters. However, the lack of objects with ages between 3 and 10 Gyr is not observed
in the field population. Finally, we used data from the literature to derive consistently the age--metallicity relationship of the bar. Simple chemical evolution models have been used to reproduce the observed age--metallicity
relationships with the purpose of investigating which mechanism has participated in the evolution of the disk and bar. We find that while the disk age--metallicity relationship is well reproduced by close-box models or models with a small degree of outflow, that of the bar is only reproduced by models with combination of infall and outflow. 
\end{abstract}


%% subject header.  Objects should be in the appropriate "individual"
%% headers (e.g. quasars: individual, stars: individual, etc.) with the
%% additional provision that the total number of headers, including each
%% individual object, not exceed six.  The \objectname{} macro, and its
%% alias \object{}, is used to mark each object.  The macro takes the object
%% name as its primary argument.  This name will appear in the paper
%% and serve as the link's anchor in the electronic edition if the name
%% is recognized by the data centers.  The macro also takes an optional
%% argument in parentheses in cases where the data center identification
%% differs from what is to be printed in the paper.



\section{Introduction}

Despite decades of work, there are still significant gaps in our
knowledge of the LMC's star formation and chemical enrichment
histories \citep{ol96}. This is motivated in
part by the vastness of its stellar populations and to our limitations
in observing sizable samples of stars. The age distribution of star
clusters is relatively well known \citep[e.g.][]{geisler97}: there
is an age interval between 10 and 3 Gyr  with almost no clusters. This so-called age-gap may
also correspond to an abundance gap \citep{ol91}, since the old
clusters are metal-poor while the young ones are relatively metal-rich. The star formation history (SFH)
of field stars is much less precisely known. Studies using {\itshape HST} data in small fields,
suggest that the SFH of the LMC disk has been more or less continuous,
with some increase in the star formation rate (SFR) in the last few Gyr
\citep[e.g.][]{Smecker-Hane02,castro01,holtzman99}. This contrasts with previous results (based on much
shallower data), which found a relatively young age (a few Gyr) for the
dominant LMC population \citep{hardy84,bertelli92,westerlund95,vallenari96}. The SFH of the
LMC can now be studied using sufficiently deep ground-based data (e.g.\ reaching the oldest main-sequence turn-off with good
photometric precision) in large areas and in different positions of the galaxy
\citep[e.g.][hereafter Paper I]{gshpz04,gshpz05}.

Less is known
about the LMC chemical enrichment history. The age--metallicity relation (AMR)
normally used for the LMC is defined by star clusters 
\citep[e.g.][]{ol91,geisler97,bica98,dirsch00}. All works based
on clusters obtain similar results.  The mean metallicity jumps from [Fe/H] $\sim -1.5$ for the oldest
clusters to [Fe/H] $\sim -0.5$ for the youngest ones. There were no studies on the field population AMR until the
last decade. \citet{dopita97}, from a study of $\alpha$-elements in planetary nebulae, obtained a
result qualitatively similar to that  found in the clusters although only ten objects were used. Later,
\citet{bica98}, using Washington photometry, and \citet{c00} and \citet{dirsch00}, using Str\"ongrem photometry,
obtained the metallicity of RGB stars in different positions of the LMC. \citet{c00} also observed 20 stars
with the infrared CaII triplet (CaT). All of them found that the age-gap observed in the clusters is not found in
the field population. More recently, \citet{c05} have obtained
stellar metallicities for almost 400 stars in the bar of the LMC, also using CaT lines. The metallicity for each star has been combined with its position in the color--magnitude diagram (CMD) to
estimate its age. The obtained AMR shows a similar behavior to that of the clusters for the oldest
population. However, while for the clusters the metallicity has increased over the last 2 Gyr, this has not happened in
the bar. Finally \citet{gratton04} measured the metallicity of about 100 bar RR Lyrae, obtaining an average
metallicity of [Fe/H] $=-1.48$.

Another point that it is necessary to investigate is the presence, or not, of an abundance gradient in the LMC.
From observations of clusters, \citet{ol91} and \citet{santos99} found no evidence for the presence of a radial
metallicity gradient in the LMC. The only evidence
for a radial metallicity gradient in the LMC cluster system was reported by \citet{kon93} 
based on six outer LMC clusters
($\geq$ 8 kpc). \citet{hill95} were the first to report evidences of a gradient in the field 
population from high-resolution spectroscopy, using a sample of nine 
stars in the bar and in the
disk. They found that the bar is on average 0.3 dex more metal-rich than the 
disk population, but the disk
stars studied are located within a radius of 2\arcdeg. Subsequent work by \citet{cioni03}, 
detected that the C/M ratio between number of asymptotic giant branch stars of spectral 
types C and M increased when moving away from the bar and within
a radius  of 6$\fdg$7. As the C/M ratio is anticorrelated with metallicity, this 
increment implies a decrease in metallicity. Finally, an outward radial
gradient of decreasing metallicity was also found by \citet{alves04} from infrared 
CMD using the 2MASS survey.

We have obtained deep photometry with the Mosaic II CCD Imager on the CTIO 4m telescope in four disk fields at
different distances from the center of the LMC (RA=5$^h$23$^m$34$\fs$5, $\delta$=-69$\arcdeg$45'22``) with a quality
similar to that obtained by {\itshape HST} in more crowded areas (see Paper I). 
The position of these fields, together with a
description of their CMDs, are presented in Paper I, 
and detailed SFHs will be published in forthcoming papers. In the present investigation we 
 focus on obtaining stellar metallicities for a significant number 
of individual RGB stars in these four fields using spectra obtained with the HYDRA spectrograph at the CTIO
4m telescope. In Section \ref{targetselection} we present our target selection. The observations and data reduction are presented in Section \ref{datareduction}. The radial velocities of the stars in our sample are obtained in Section \ref{radialvelocities}. In Section \ref{cat} we discuss the calculation of the CaT equivalent widths and the determination of metallicities. Section \ref{agedetermination} presents the method used to derive ages for each star by combining the information on their metallicity and position on the CMD. The analysis of the data is presented in Section \ref{analysis}, where the possible presence of a kinematically hot halo is discussed and the AMR and the SFH in each of our fields are described and compared. In the final section, the AMRs are compared to theoretical models and conclusions are drawn about the chemical evolution of the LMC.
 
\section{Target Selection\label{targetselection}}

Our starting point are deep CMDs of four $36'\times 36'$ fields located $\sim$3$\arcdeg$,
5$\arcdeg$, 6$\arcdeg$ and 8$\arcdeg$ from the LMC center, which are presented and discussed in Paper I. All fields are located north of the
LMC bar, and their
CMDs reach the
oldest main-sequence turn-offs. A clear gradient exists in the amount of intermediate-age
and young populations from the inner to the outermost fields, in the sense that the 
young population is more prominent in the inner parts
(see Paper I for a more detailed discussion). 

In each field we have selected the stars to be observed with HYDRA in two windows in the CMD, which are plotted in Figure \ref{dcmbox}. The bluest color limit
has been selected to avoid upper red-clump and red supergiant stars and the lower magnitude limit to avoid including red-clump stars in the spectroscopic sample. The blue edge of the window is sufficiently blue to ensure that no metal 
poor
stars would be excluded (the position of the blue ridge corresponds
approximately with the position of $\sim$2 Gyr old stars with [Fe/H]=-3).
Similarly, the opposite edge is sufficiently red so that even old stars with
solar metalicities, or even higher, fit within the box. These selected stars
have been ordered from the brightest to the faintest ones, with no color restriction. This list has been
used as input for the configuration task of the instrument, which tries to optimize the number of allocated fibers. The first stars in the list, the brightest ones, have priority over the others.

\section{Observations and Data Reduction\label{datareduction}}

The observations were carried out in two different runs, in December 2002 and in
January 2005, at the CTIO 4 m telescope with the HYDRA multifiber spectrograph. We used
the KPGLD grating, which provides a central wavelength of
8500 \AA, and a OG590 order-blocking filter. The physical pixels were binned 2 $\times$ 1 in the spectral direction yielding a
dispersion of 0.9 \AA /pix. HYDRA has a field of view of about 40 arcmin,
which on account of the high density of stars in our fields, allowed us to observe more than
100 stars in each configuration, with the 140 available fibers.  Because of technical problems and bad weather,
on our first run we were able to observe only two
configurations located at 5$\arcdeg$ and 8$\arcdeg$ respectively. In the more successful second run,
one fiber configuration in each field was observed except for the closest one, where
we observed two configurations. Stars in two globular clusters, NGC 4590 and NGC 3201, were also observed. Those in the first were used as radial velocity standards, but also to 
compare the equivalent widths determined through the present set-up with those setups
used for the cluster observations reported in 
\citet[][hereafter Paper II]{carrera06}. Since the differences were 
negligible (see Figure 1 of paper II), the second cluster was included 
in the calibration of the CaT as metallicity indicator. We concluded that the LMC equivalent widths obtained with the present observational setup could be directly 
transformed into [Fe/H] using the calibration in paper II. Equivalent widths, magnitudes and radial velocities for the 702 stars observed in the LMC are listed in Table \ref{starobs}.

The data reduction was done following the procedure described in {\sl Knut Hydra notes}\footnote{http://www.ctio.noao.edu/spectrographs/hydra/hydra-knutnotes.html}.
First, cosmic rays were removed from the images using the 
IRAF\footnote{IRAF is distributed by the National Optical Astronomy
Observatory, which is operated by the Association of Universities for 
Research in Astronomy, Inc., under cooperative agreement with the
National Science Foundation.} Laplacian edge-detection
routine \citep{dokkun01}. All images were then bias- and 
overscan-subtracted, and trimmed. Flat exposures were taken during the day with a diffusing screen installed in the spectrograph to obtain
the so-called "milk-flat", which is used to correct the effect of the different sensitivities among pixels. It was applied to all images
using \textsl{ccdproc}. The spectra were extracted, flat-field
corrected (see bellow), and calibrated in wavelength using \textsl{dohydra}, a program developed
specifically to reduce the data obtained with this spectrograph. The lamp flats acquired
at the beginning of each configuration were used to define and trace each aperture. The sky flats were used to
eliminate the residuals after the flat-field correction with the milk-flat. Arcs, obtained before and after each
configuration, were used for
wavelength calibration.

\textsl{Dohydra} can also perform sky subtraction. However, we noticed that after sky
subtraction, there remained  significant sky line residues.  We therefore
 developed our own procedure to remove the contribution
of the sky lines from the object spectra. For a given configuration, we obtain an
average high signal--to--noise ratio (S/N) sky spectrum from all fibers placed on the sky. 
Before subtracting this averaged sky from each
star spectrum, we need to know the relation between the intensity of the sky in 
each fiber (which varies from fiber to fiber owing
to the different fiber responses) and the average sky. This relation is used as 
a weight (which may depend on wavelength) that
multiplies the average sky spectrum before subtracting it from each star. Our task 
minimizes the sky line residuals over the whole spectral region considered and 
allows very accurate removal of the sky emission lines. An example of raw, sky and final object
 is shown in Figure \ref{sky}.

We obtained between three and four exposures of 2700 s in each configuration. 
To minimize the contribution of
residuals arising from cosmic rays and bad pixels, all spectra of the same 
object were combined. Finally, each object spectrum was normalized with the
\textsl{continuum} task by fitting a polynomial, making sure to 
exclude all absorption lines in our wavelength range, such as the CaT lines themselves.
 
\section{Radial Velocities\label{radialvelocities}}

Our main purpose in obtaining radial velocities for our program stars is to 
reject possible foreground objects. However, radial velocities
can provide more useful information, as discussed in depth in Section \ref{kinematics}. 
The radial velocities were calculated with the \textsl{fxcor} task
in IRAF, which performs the cross-correlation described by \citet{td79} 
between the spectra of the target and selected templates of known radial
velocities. As templates, we used three stars in the globular cluster NGC 4590 (M68) 
with radial velocities obtained by \citet{geisler95}.
The final radial velocities for all the stars in our sample were obtained as the average 
of the velocities obtained with the three templates, weighted by the width of the corresponding correlation
peaks. Radial velocity histograms for each field are plotted in Figure \ref{vel_dist}. 
A Gaussian function was fitted in order to obtain
the central peak and the $\sigma$ of each
distribution (values listed in Table \ref{radialvelocity}). The velocity
distribution of the field  located 3\arcdeg\ from the center shows two peaks, which are marked in Figure
\ref{vel_dist}. Stars with radial velocity in the range
170 $\leq V_r\leq$ 380 km s$^{-1}$ \citep{zhao03} are considered  LMC members. Only about 20 stars in each
configuration of the instrument have been excluded based on their radial velocity. These results are discussed in depth in Section \ref{kinematics}.

\section{CaT Equivalent Widths and Metallicity Determination\label{cat}}

The metallicity of the RGB stars is obtained following the procedure described in Paper II.
In short, the equivalent width is the line area normalized to the continuum. The continuum is
calculated from the linear fit to the mean value of each bandpass, defined to obtain the continuum position. We
have used the continuum and line bandpasses defined by \citet{cen01}, which are listed in Table \ref{bandastable}. The
equivalent widths are calculated from profile fitting using a Gaussian plus
a Lorentzian. This combination provides the best fit to line core and wings as discussed in Paper II. The CaT index is defined as the sum of the equivalent widths of the three CaT lines,
denoted as $\Sigma Ca$. The $\Sigma Ca$ calculated for each observed star and its uncertainty are given in
Table \ref{starobs}, together with the star magnitude and radial velocity. The reduced equivalent width,
$W'_I$, for each star has been calculated as the value of $\Sigma Ca$ at M$_I$=0, using the slope obtained in Paper II for the calibration
clusters in the M$_I$--$\Sigma Ca$ plane ($\beta_I$ = $-0.611$ \AA mag$^{-1}$).

In Paper II we obtained relationships between the reduced equivalent widths, $W'_V$ and $W'_I$, and
metallicity, on the
\citet{zw84}, \citet[hereafter CG97]{cg97} and \citet{ki03} scales. In this case, we will only use the
relationships obtained on
the CG97 metallicity scale, because it is the only one in which the metallicities of our open and globular calibration clusters have been obtained in a homogeneous way from high-resolution spectroscopy.

In order to verify that the $V$ and $I$ relationships give similar results, we have applied both to obtain the
metallicities of the stars observed in the field located at 5\arcdeg. The differences are on average $-$0.01
dex, and never larger than 0.15 dex for stars with $(V-I)\leq$ 2.5. For stars with colors larger than this, the
differences are on average 0.2 dex and never get
above 0.35 dex. The calibration stars used in Paper II have $(V-I)<2.5$, so the
metallicities calculated for the reddest stars are necessarily extrapolations of the 
relationships. On average, the relationship
based on M$_I$ yields metallicities slightly lower than those obtained from M$_V$, 
but always within the uncertainties. In what follows we will use
the relationship based on M$_I$, because: i) the photometric accuracy in $I$
is slightly better than in $V$, especially for the reddest
stars on the RGB; ii) the RGB is better resolved in the $I$ band; and iii) from a theoretical
point of view \citep{pont04},
it is likely that the relationship based on M$_I$ is less sensitive to age. In short, then, the metallicity for each star is given by:

\begin{equation}\label{metaleq}
[Fe/H]_{CG97}=-2.95+0.38\Sigma Ca+0.23M_I
\end{equation}

Binarity could affect the derived metallicity only in the unlikely case of the two stars in the system being RGB stars of a very similar mass. In this case the system would be $\sim$0.7 mag brighter than a single star. As the binarity might not affect the equivalent width of the CaT lines, the derived metallicity would be lower by $\sim$0.4 dex than the actual metallicity.

Figure \ref{misigmaca} shows the position of LMC radial velocity member stars in the 
M$_I$--$\Sigma$Ca plane for our
four fields, from the innermost one (top left) to the outermost field (bottom right). The star with $\Sigma Ca$ = 11 and M$_I$ = $-$2 in the field
situated at 5$\arcdeg$\@ has a well measured $\Sigma Ca$ and radial velocity 
V$_r$ = 264 km s$^{-1}$ (the mean V$_r$ of the sample is 278 $\pm$ 20 km
s$^{-1}$), so it seems to be an LMC RGB star with unusually large metallicity. However, 
a reliable metallicity cannot be obtained with the
present calibration, which is only valid for [Fe/H] $\leq$ +0.47. The extrapolation of our 
calibration gives [Fe/H] = +0.72 $\pm$ 0.4 for this
star.

The metallicity distribution of each field is shown in Figure
\ref{histo_fields} and will be discussed \emph{in extenso} in Section \ref{analisismetallicitydistribution}. In
Figure \ref{dcm_feh} the observed stars have been plotted  in
the color--magnitude diagram  employing different symbols as a function of their metallicity. 
It is important to highlight that there is no correlation between the
metallicity of the stars and their position in the color--magnitude diagram,
as was also found in other galaxies \citep[e.g.][]{pont04,c05,koch05}. This shows once more that the metallicities derived
from the position of the stars in the RGB are unreliable in systems with complex SFHs.

\section{Determination of stellar ages\label{agedetermination}}

For a given age, the position of the RGB on the color--magnitude diagram depends mainly on metallicity
in the sense that more metal-rich stars are redder than more metal-poor ones. At the same time, for a fixed
metallicity, older stars are also
redder than younger ones. If we take into account the logical assumption that younger stars 
are also more metal-rich, age may partly counteract the effect of
metallicity. The combination of both effects is the well known RGB age--metallicity 
degeneracy, which sets a limit on the amount of information that can be retrieved from the
RGB using only photometric data. However, if stellar metallicities are obtained independently from another source, 
such as spectroscopy, we should in principle be able to break the age--metallicity
degeneracy. \citet{pont04} and \citet{c05} have derived stellar ages from isochrones 
for stars whose metallicity had been previously
obtained from low-resolution spectroscopy. In our case, instead of
directly comparing  the position of the star in the CMDs with 
isochrones, we have obtained a relation for the age as a
function of color, magnitude and metallicity from a synthetic CMD. This allows to easily obtain ages for the large number of stars in our sample. 

The reference synthetic CMD has been computed using IAC-STAR\footnote{available on the Web
at http://iac-star.iac.es} \citep{aparicio04}. Its main input parameters are the SFR and
the chemical enrichment law, as a function of time. Since we are interested in a general relation, we have chosen a
constant SFR for the whole range of ages (13 $\geq$ (Age/Gyr) $\geq$ 0) and a chemical enrichment law such that a star of any age
can have any metallicity in the range $-2.5\leq$ [Fe/H] $\leq$ +0.5. Both the constant SFR and the chemical enrichment
law ensure that there is no age or metallicity bias. We have computed two
synthetic CMDs, one using the BaSTI
stellar evolution library \citep{pie04} and another  using the Padova library \citep{girardi02}
 respectively. The ranges of ages and metallicities
covered, as well as other important features of each library are listed in Table 1 of \citet{gallartza05}.

\begin{figure}
\epsscale{1}
\plotone{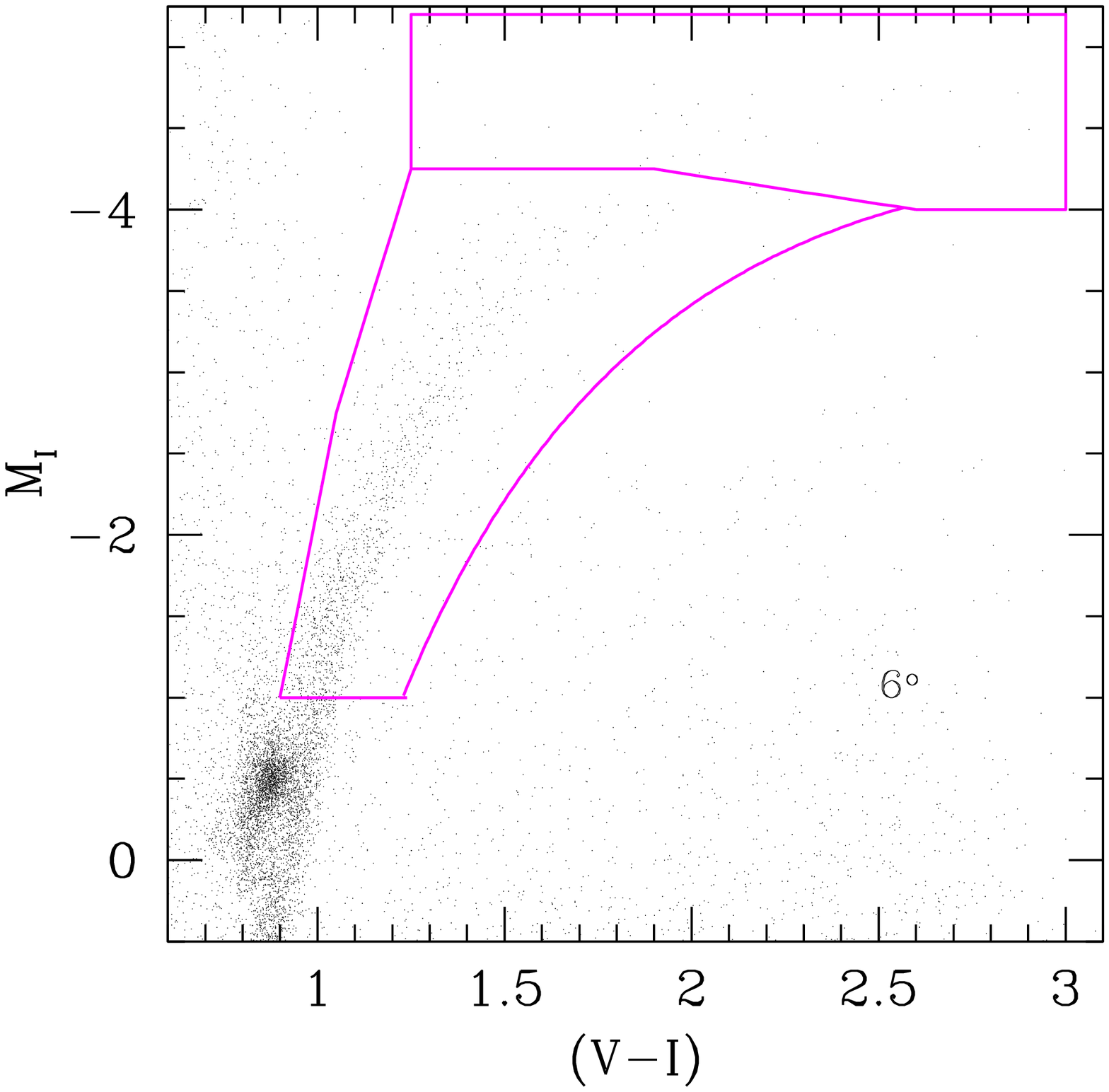}
\caption{CMD of the RGB region of the field situated at 6\arcdeg\@ from the LMC center. The regions used to
select stars to be observed spectroscopically are shown.\label{dcmbox}}
\end{figure}

\begin{figure}
\epsscale{1}
\plotone{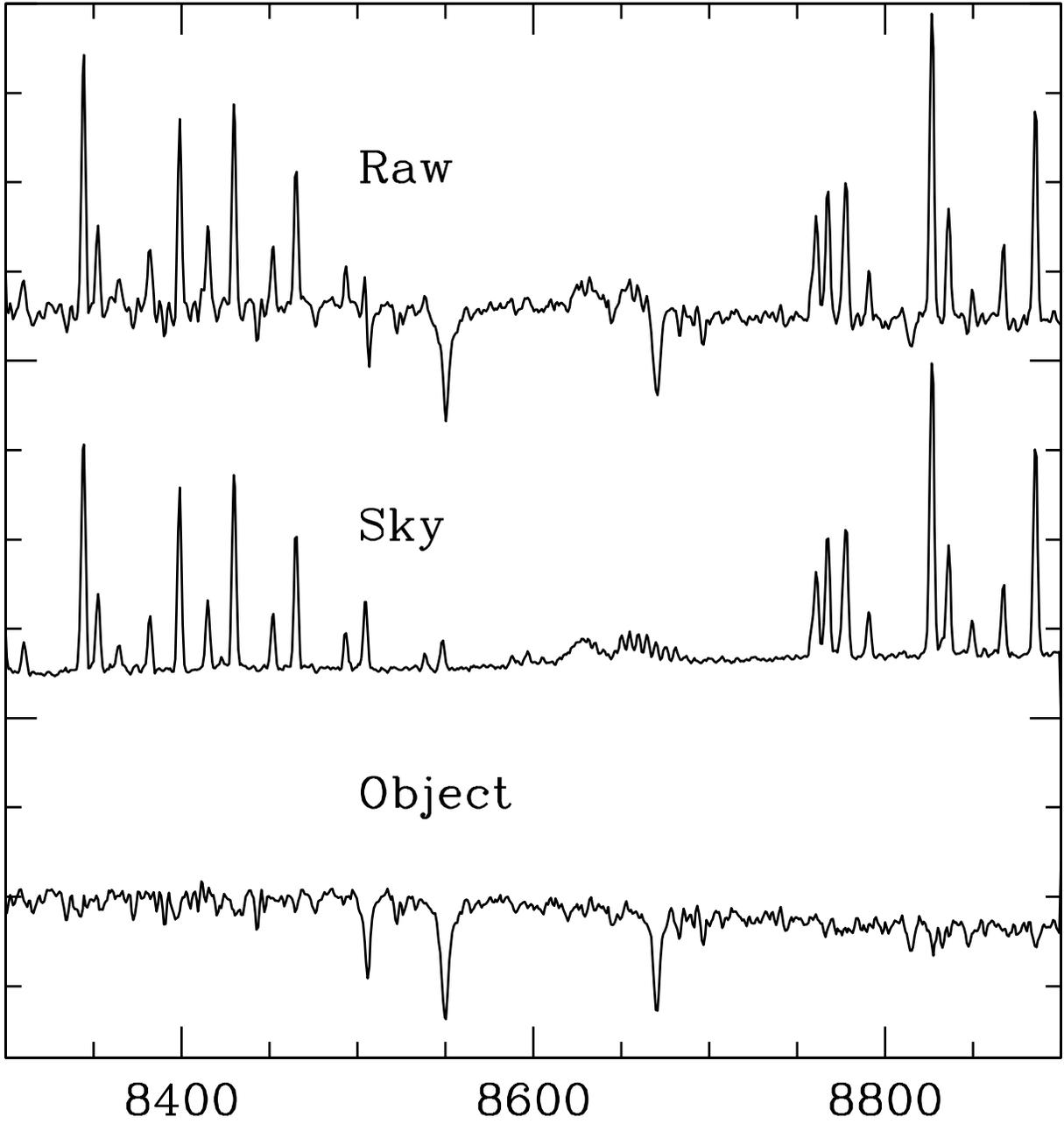}
\caption{Example of sky subtraction. Top: raw spectrum. Middle: sky spectrum obtained
 from  a combination of fibers located on the
sky. Bottom: The final spectrum after  sky subtraction.\label{sky}}
\end{figure}

\begin{figure}
\epsscale{1}
\plotone{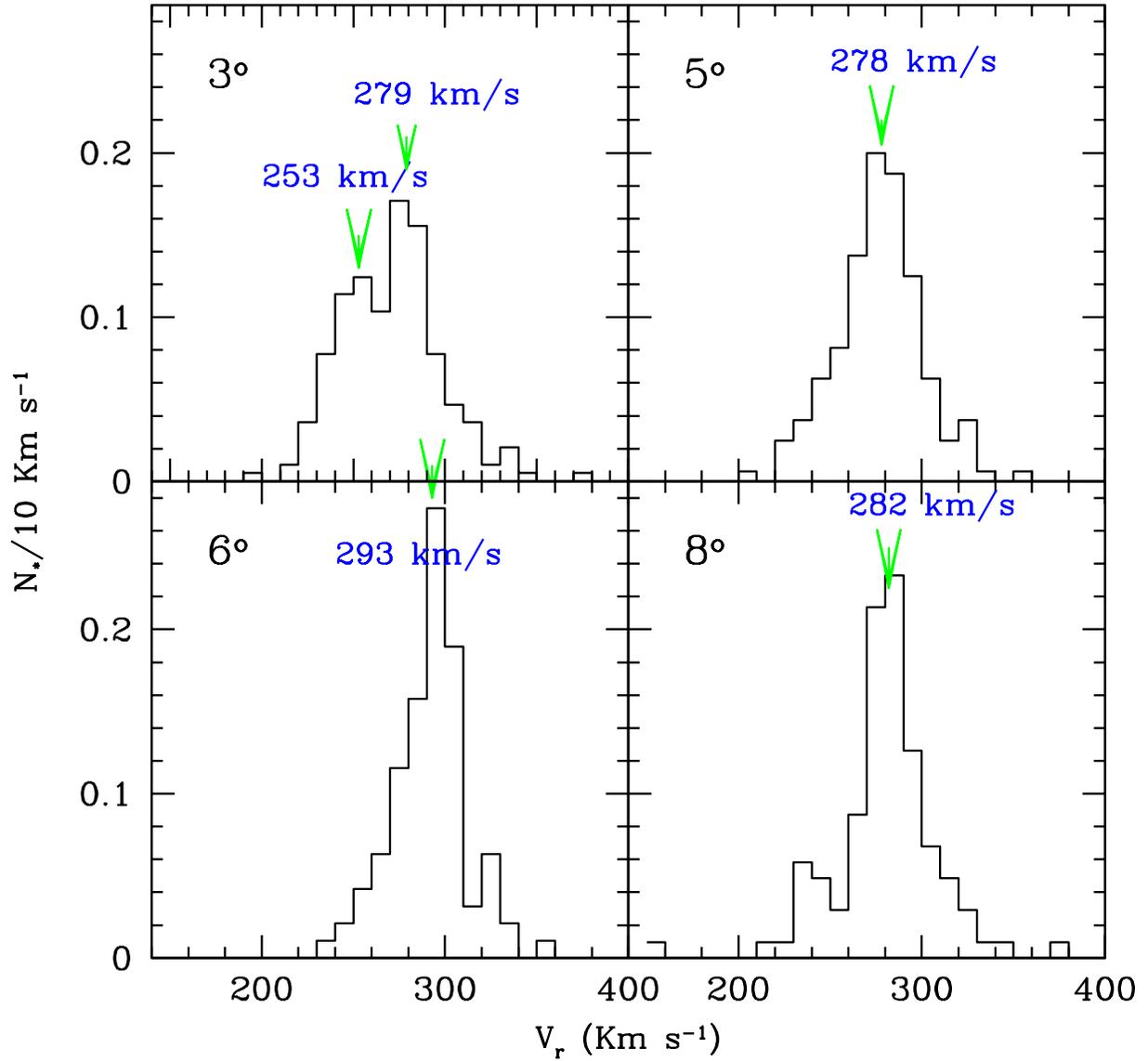}
\caption{Radial velocity distributions of our four LMC fields. The mean value of each distribution is shown in
each panel, and its position is marked with an arrow. Note that there are two peaks in the field situated at
3\arcdeg.\label{vel_dist}}
\end{figure}

\begin{figure}
\epsscale{1}
\plotone{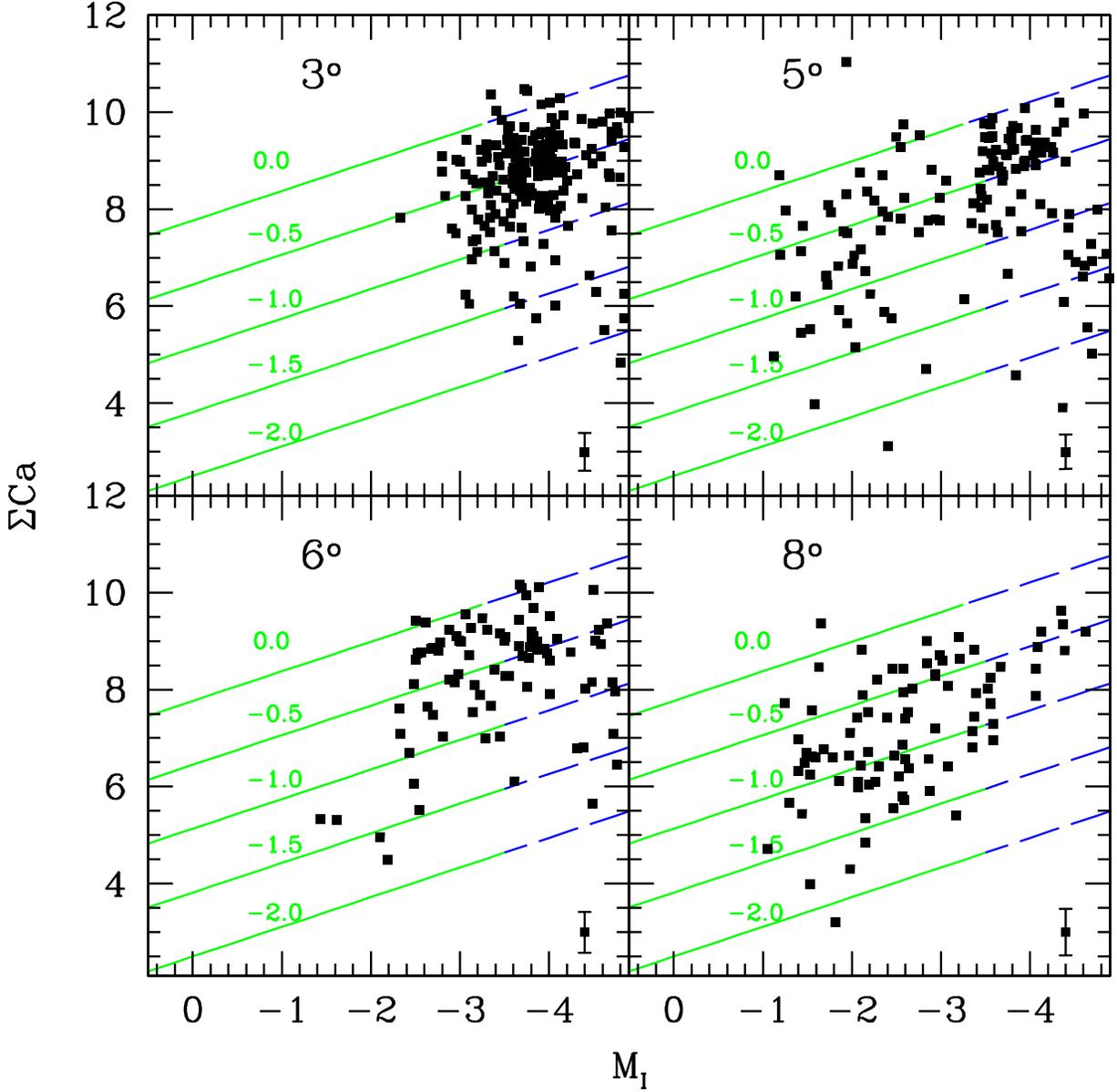}
\caption{Position in the M$_I$--$\Sigma$Ca plane of LMC stars for which membership has been confirmed from
their radial velocity. The typical $\Sigma Ca$ error is shown in the bottom right corner of each panel.
Isometallicity lines have been plotted
for reference. The
solid part of each line is the magnitude interval covered by the cluster stars used for the calibration
(see Paper II). The dashed part is the region in which the calibration is
extrapolated.\label{misigmaca}}
\end{figure}

\begin{figure}
\epsscale{1}
\plotone{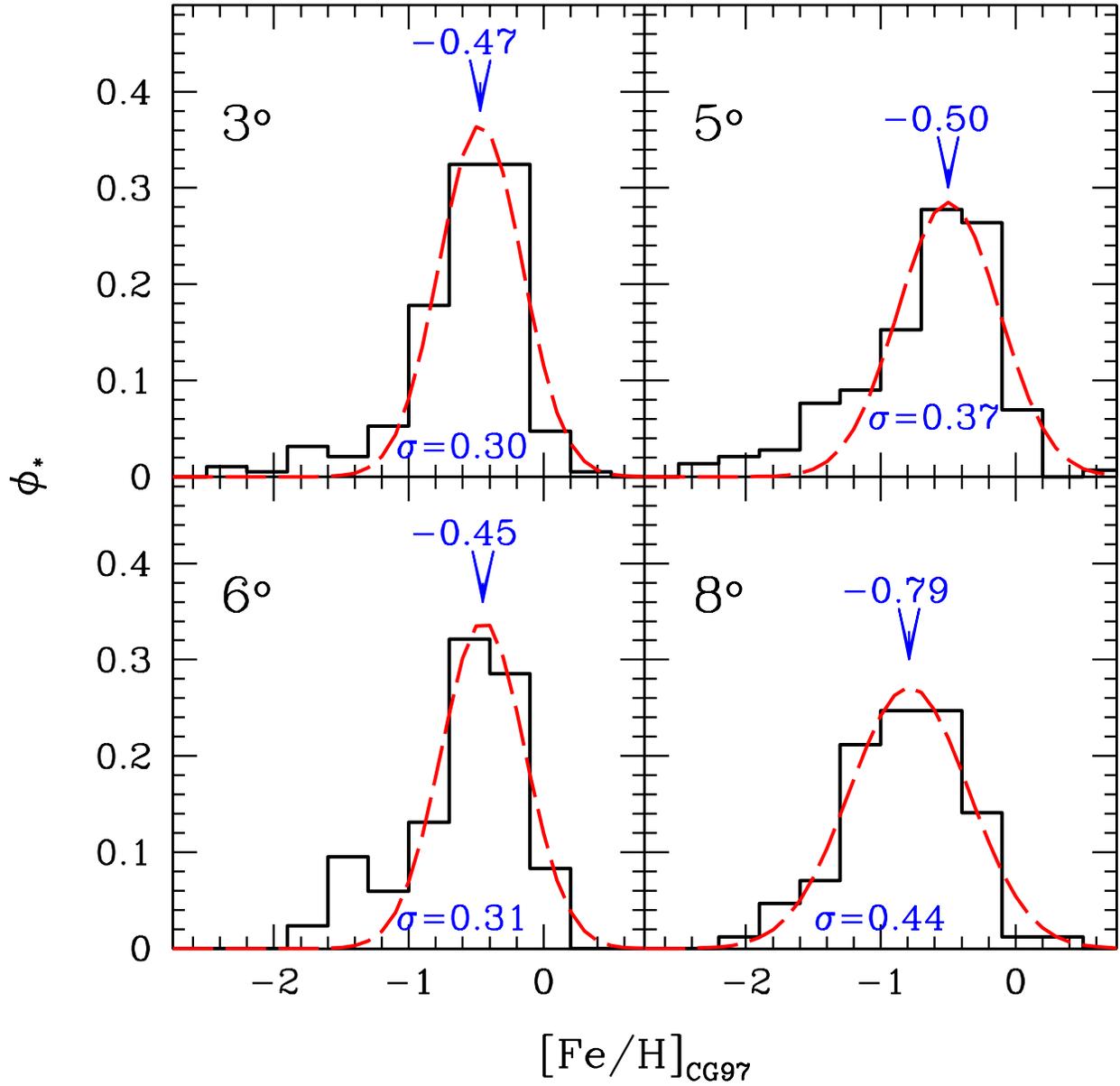}
\caption{Metallicity distributions for the four fields in our sample. A Gaussian has been fitted to 
each distribution in order to obtain its
mean and  dispersion. The values obtained are shown in each panel.\label{histo_fields}}
\end{figure}

\begin{figure}
\epsscale{1}
\plotone{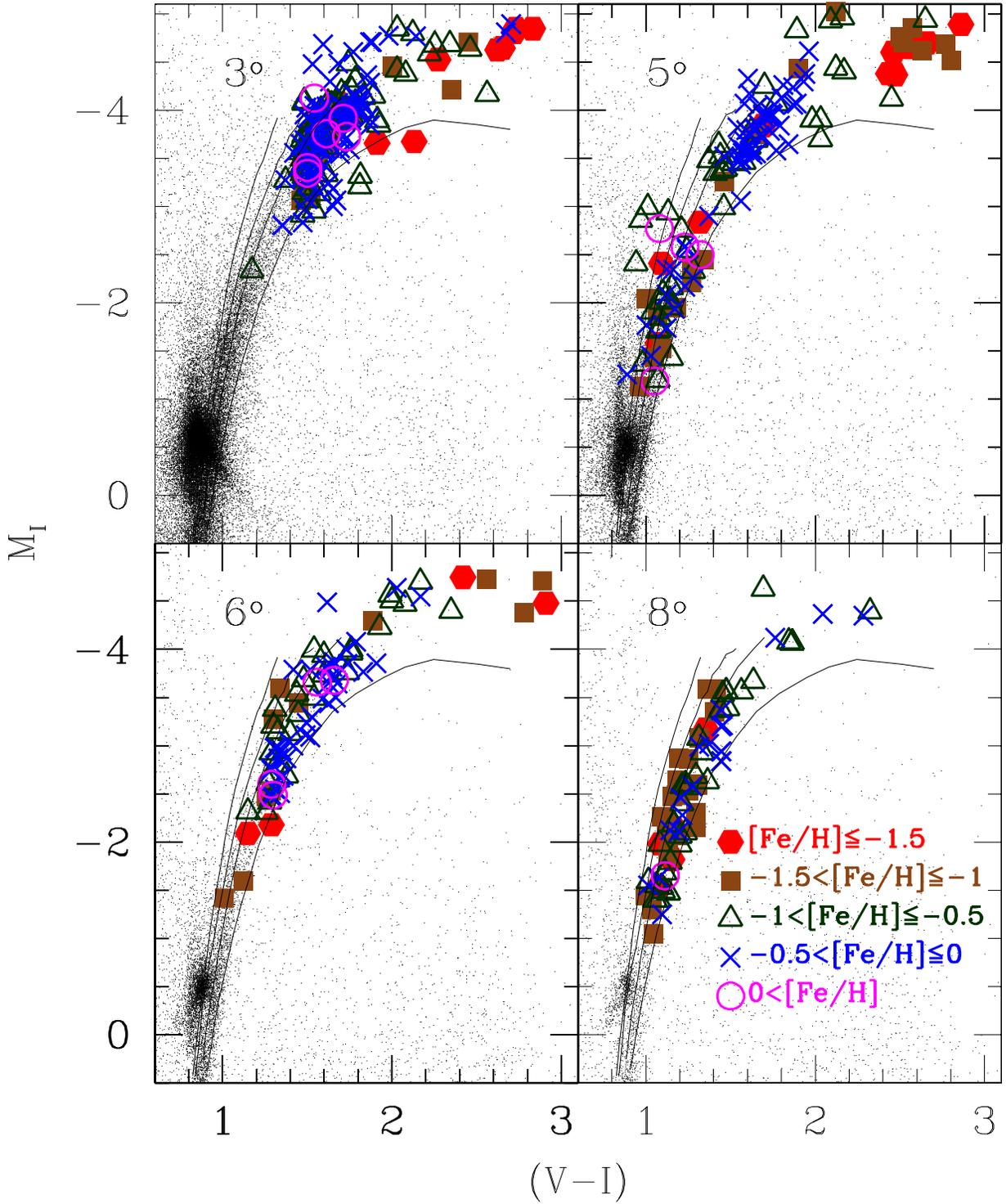}
\caption{Position of the observed star in the
color-magnitude diagram. Different metallicity intervals are represented by different symbols, as indicated in the label. Solid lines are
the fiducial sequences of four globular clusters (from right to left: NGC 104: [Fe/H] = $-0.78$; NGC 5904:
[Fe/H] =$-1.12$; NGC 5272: [Fe/H] = $-1.34$ and NGC 6341: [Fe/H] = $-2.16$).\label{dcm_feh}}
\end{figure}

Since the uncertainty of the age assigned by stellar evolution models to bright AGB stars is large, we will only compute ages for RGB and AGB stars fainter than the RGB tip (M$_I\sim$-4 in Figure \ref{dcmbox})
. We have chosen the synthetic stars in the box below the tip of the RGB shown in
Figure \ref{dcmbox}. We have computed a polynomial relationship which gives
the age as a function of [Fe/H] $\equiv$\   log(Z/Z$_\odot$), (V-I) and M$_V$. In order to 
minimize the $\sigma$ and to improve the correlation coefficient of the relation, 
different linear, quadratic and cubic terms of each observed magnitude have been added. When the addition of
a term did not improve the relation we rejected it. The best combination of these
parameters is in the form of Equation \ref{rela} and the values obtained for each stellar evolution
library are shown in Table \ref{tableage}. Note that each stellar evolution library has
 its own age scale, and some differences might exist among them.

\begin{equation}\label{rela}
log(age)=a+b(V-I)+cM_V+d[Fe/H]+f(V-I)^2+g[Fe/H]^2+h(V-I)^3
\end{equation}
 
\begin{figure}
\epsscale{1}
\plotone{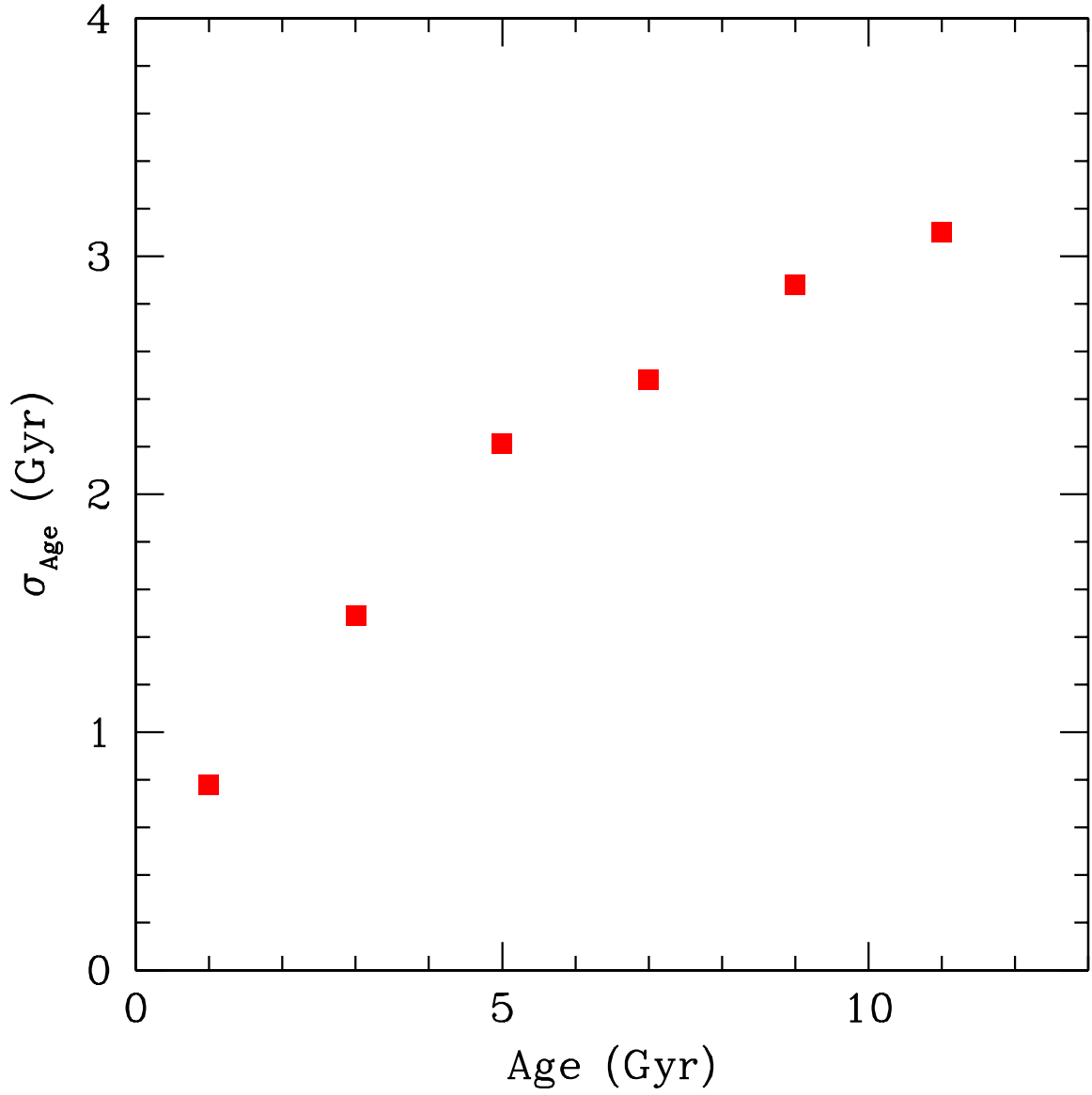}
\caption{Uncertainty in the age calculated using Equation \ref{rela} obtained from the BaSTI stellar evolution
models. See text for details.\label{erroredad}}
\end{figure}

The uncertainties of each term on Equation \ref{rela} are given in Table \ref{tableage}. However, the way in which they
propagate into the error of the age is complex. For this reason a Monte Carlo test has been performed to
estimate age errors. The synthetic stellar population computed from BaSTI stellar evolution models has been
used to this purpose. That based on the Padua library would have produced similar results. The test consists
in computing, for each synthetic star, several age values for stochastically varying $[Fe/H]$, $(V-I)$ and $M_{V}$ according  to a gaussian probability
distribution of the corresponding $\sigma$ ($\sigma_{[Fe/H]}\sim$0.15 dex; $\sigma_{(V-I)}\sim$0.001 and
$\sigma_{M_V}\sim$ 0.001). The $\sigma$ value of the obtained ages provide an estimation of the age error when
Equation \ref{rela} is used. The values obtained for different age intervals are shown in Figure \ref{erroredad}. The error increases for older ages.

We have also investigated the accuracy with which the age of a real stellar population is measured. The
stars in open and globular clusters in Paper II, which ages are independently
known, are used for this purpose. We adopt
 the ages estimated by \citet{sw02} and \citet{swp04}, which are in a common scale, for a globular and open clusters, respectively. They are listed in Table 1
of Paper II. 

\begin{figure}
\epsscale{0.6}
\plotone{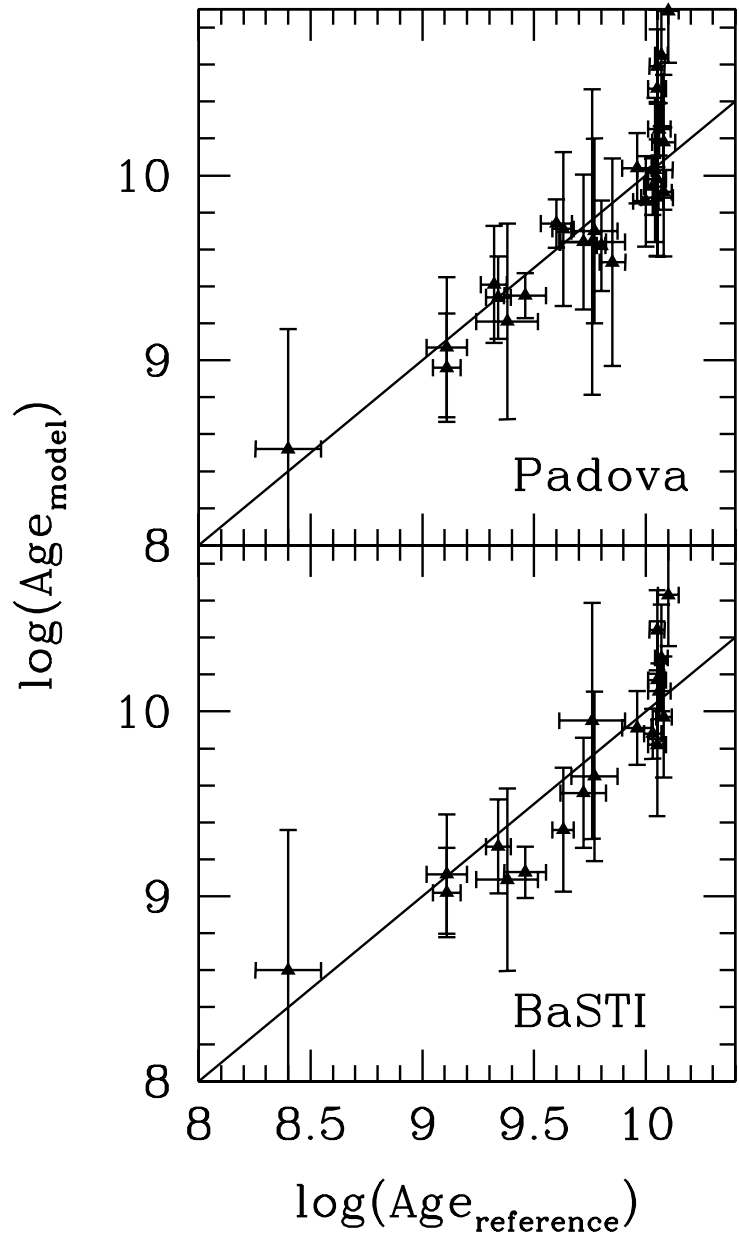}
\caption{Ages derived for the cluster sample in Paper II, using Equation \ref{rela} and BaSTI and Padova
models versus the reference ages. The solid lines are the one-to-one relation. 
\label{cluster}}
\end{figure}

The age of each
cluster star has been estimated from Equation \ref{rela} assuming the distance modulus and reddening listed in Paper II. The metallicity of each
star has been calculated following the same procedure as for the LMC stars 
(see Section \ref{cat}). The age of each cluster has been obtained as the mean age derived from 
all the stars in each cluster and the quoted uncertainty is the standard deviation of this mean. Ages derived for each cluster from the BaSTI and Padova
relationships in Table \ref{tableage}, versus the reference values for each
cluster, have been plotted in Figure \ref{cluster}. The relationships saturate for ages older than 10 Gyr
because differences of 1 or 2 Gyr
produce only negligible variations of position in the 
CMD. In both cases, the 
age of clusters younger than 10 Gyr are well reproduced. Differences between the age derived with each model 
and the reference value are within the error bars, with the exception of cluster NGC 6819 for the BaSTI relationship. The large errorbar of the youngest cluster, NGC 6705, is due to the fact that the stars observed in this cluster are fainter that the synthetic stars used to derive the relationships. Therefore, we are extrapolating the relations to obtain the age of this cluster. As both models produce similar values, we adopted for simplicity the relationships derived from the BaSTI stellar library \citep{pie04}.

\section{Analysis\label{analysis}}

\subsection{Stellar kinematics of the LMC\label{kinematics}} 

\begin{figure}
\epsscale{1}
\plotone{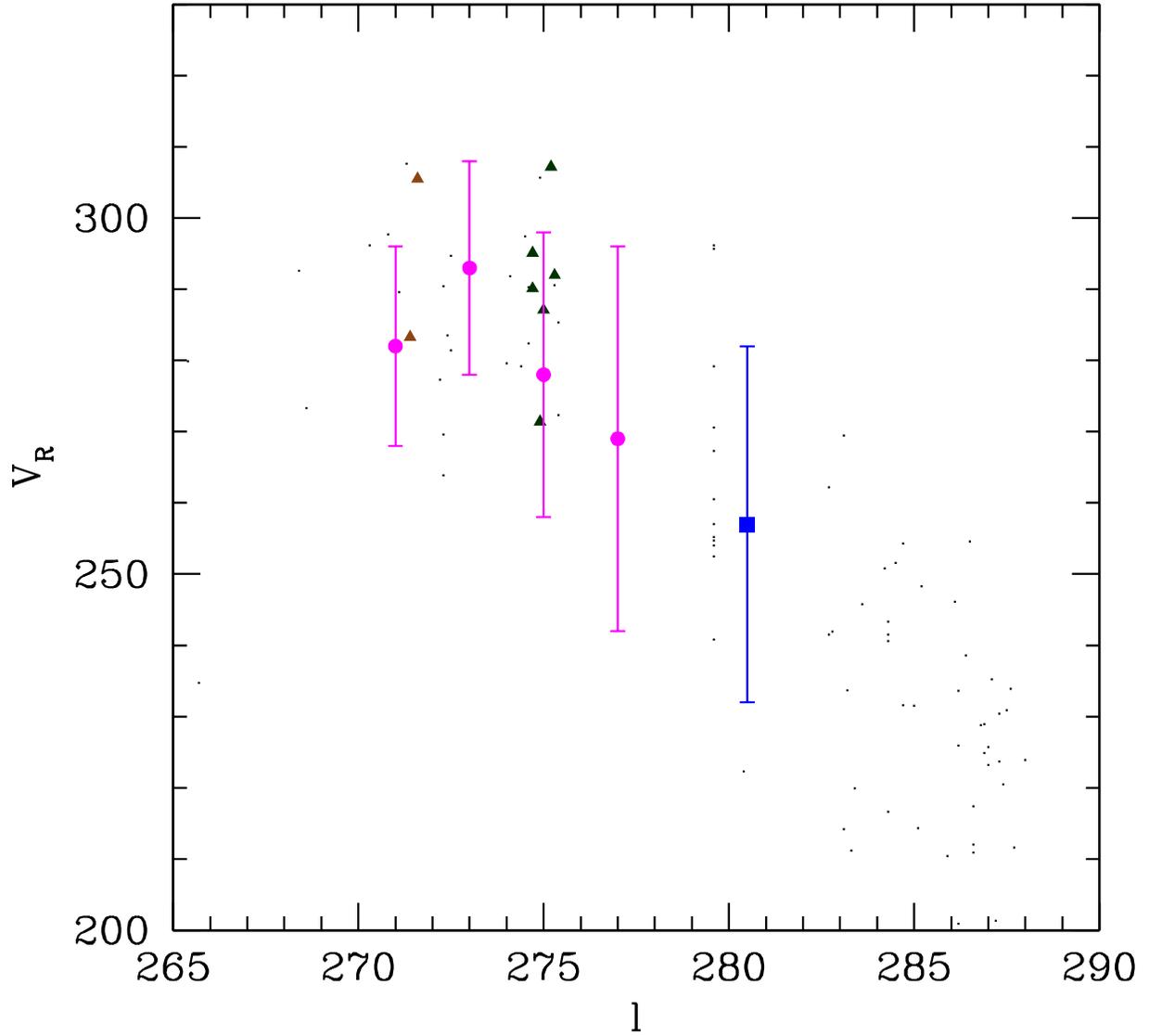}
\caption{Radial velocities for carbon stars from \citet{kunkel97} in 
the band $-36\arcdeg \leq b \leq -33\arcdeg$. The triangles
are the carbon stars in our fields at 
5\arcdeg~and 8\arcdeg. The filled square represents the
data from \citet{c05} for the bar. The filled dots are our data listed in Table \ref{radialvelocity}. 
The error bars represent the dispersions of the
velocity distributions.\label{kunkel}}
\end{figure}

In Figure \ref{kunkel} we have plotted the mean of the velocity distribution for each field 
\citep[including the][ bar field]{c05} as a function of
its Galactic longitude. The mean velocity changes from field to field due to the disk rotation
of the LMC. The radial velocities of carbon stars
derived by \citet{kunkel97} have also been plotted. Our data are in good agreement with their result,
 which is consistent with the presence of 
a rotational disk.

If a classical Milky Way-like halo existed in the LMC (i.e. old, metal-poor and with high velocity dispersion), a dependency between the velocity dispersion and metallicity/age of the stars in our sample might be observed. The first evidence of the
presence of a kinetically hot spheroidal population was reported by \citet{hughes91}. They found that the LMC
long period variables, related to an old stellar population, have a high velocity dispersion (33 km
s$^{-1}$),
with a low rotational component. \citet{miniti03}
 measured the kinematics of 43 RR Lyrae stars in the inner regions of the LMC and 
 found that the velocity dispersion of these stars
is
53 $\pm$ 10 km s$^{-1}$, which they associated with the presence of a kinematically hot halo populated by old 
metal-poor stars. Some studies of the intermediate-age and old populations
have found that the velocity dispersion increases with age 
\citep[e.g.][]{hughes91, schommer92, graff00}. In fact \citet{graff00}, using C stars, have found the stars of the disk
belong to two populations: a young disk population containing 20\% of stars with a velocity dispersion of 8 km
s$^{-1}$, and an old disk population containing the remaining stars with a velocity dispersion of 22 km
s$^{-1}$.

We can check the presence of a hot halo with our sample stars. Assuming that stars with similar metallicities are of similar ages (see below), a possible dependency of velocity dispersion with metallicity could indicate whether different stellar populations have different kinematics.

\begin{figure}
\epsscale{1}
\plotone{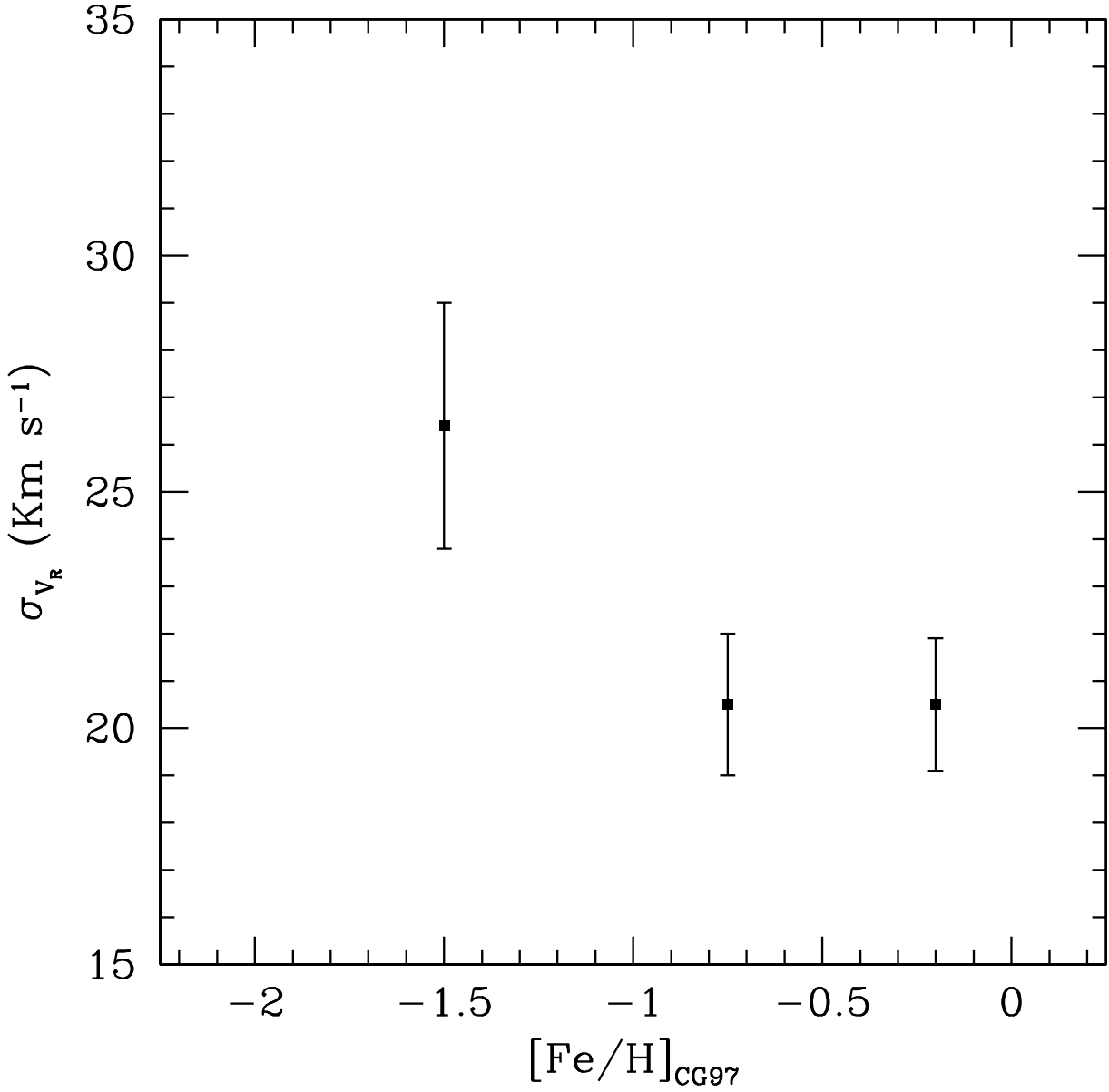}
\caption{Radial velocity dispersion of the disk stars as a function of metallicity. 
Although the velocity dispersion increases from metal-rich to metal-poor stars, this increment is not
clear enough to conclude that a metal-poor stellar halo is present.\label{vrmetal}}
\end{figure}

The procedure has been the follow up. The mean velocity in each field has been subtracted from the radial velocity of each star to eliminate the rotational component of the disk. The total metallicity range has been divided into three bins. The velocity dispersion of stars in each metallicity bin is listed
in Table \ref{vrmetaltable} and plotted in Figure \ref{vrmetal}. The velocity dispersion for old and
metal-poor stars, 26.4 km s$^{-1}$, is slightly smaller than the value found by \citet{hughes91} for the old long period variables
($\sigma$=33 km s$^{-1}$) and significantly smaller than the dispersion found in RR Lyrae stars
\citep[$\sigma\sim$53 km s$^{-1}$,][]{miniti03}. \citet{c05} found a velocity dispersion of $\sigma$ = 40.8 km
s$^{-1}$ for the most metal-poor stars in the bar, which is also higher than the value found here. The metal-poor stars in our sample seem to be members of the thick disk instead of the halo. In
short, from our data we have found no clear evidence for the presence of a kinematically hot halo populated by old stars.

\subsection{Metallicity distribution\label{analisismetallicitydistribution}}

The metallicity distribution for each field is shown in Figure \ref{histo_fields}. We have fitted a Gaussian to
each one in order to obtain its mean value and  dispersion (Table \ref{metallicitybin}). The mean value is constant at [Fe/H]$\sim$-0.5 dex up to
8$\arcdeg$, where the mean metallicity decreases by about a factor of two ([Fe/H]=-0.8 dex). All
metallicity distributions are similar. They show a clear peak with a tail toward  low metallicities. We 
complement our sample with the results by \citet{c05}, who derived 
stellar metallicities in the bar in a similar way as here. 
As demonstrated in Paper II, their CaT index is equivalent to ours. Using their
measurements of the $\Sigma Ca$ we can calculate the metallicities from Equation \ref{metaleq}.
We have also fitted a Gaussian to the resulting
bar metallicity
 distribution to obtain its mean value. On average, the bar is slightly more 
metal-rich than the inne disk [Fe/H]$\sim$-0.4 dex).
  
\subsection{Age--metallicity relationships\label{agemetallicity}}

To understand the nature of the observed metallicity distributions, and to join insight on SFR
and the chemical enrichment history of the LMC fields, we have estimated the age of each star in our sample
following the procedure described in Section \ref{agedetermination}. These individual age determinations have a much 
larger uncertainty than the metallicity calculations. However, they are still 
useful because we are interested in the general trend rather
than in obtaining precise values for individual stars.

We assumed that the oldest stars 
have the same age as the oldest cluster in our
galaxy, for which we adopt 12.9 Gyr \citep[NGC 6426,][]{sw02}. This agrees with a 13.7$\pm$0.2 Gyr
old Universe \citep{spergel03} where the
first stars formed about 1 Gyr after the Big Bang. Another important point is the youngest age that we can find in our sample. According to stellar evolution models \citep{pie04}, we do not expect
to find stars younger than 0.8 Gyr in the
region of the RGB where we selected our objects. However, application of Equation \ref{rela} may result in ages younger
than this value. To avoid this contradiction, and taking into account that the 
age determination uncertainty for the younger stars is about 1 Gyr, we assigned an age of 0.8 Gyr to those
stars for which Equation \ref{rela} gives younger values. Finally, as the relationships used to estimate
the age have been calculated for stars below the tip of the RGB, only stars fainter than this point have been
used. 

\begin{figure}
\epsscale{1}
\plotone{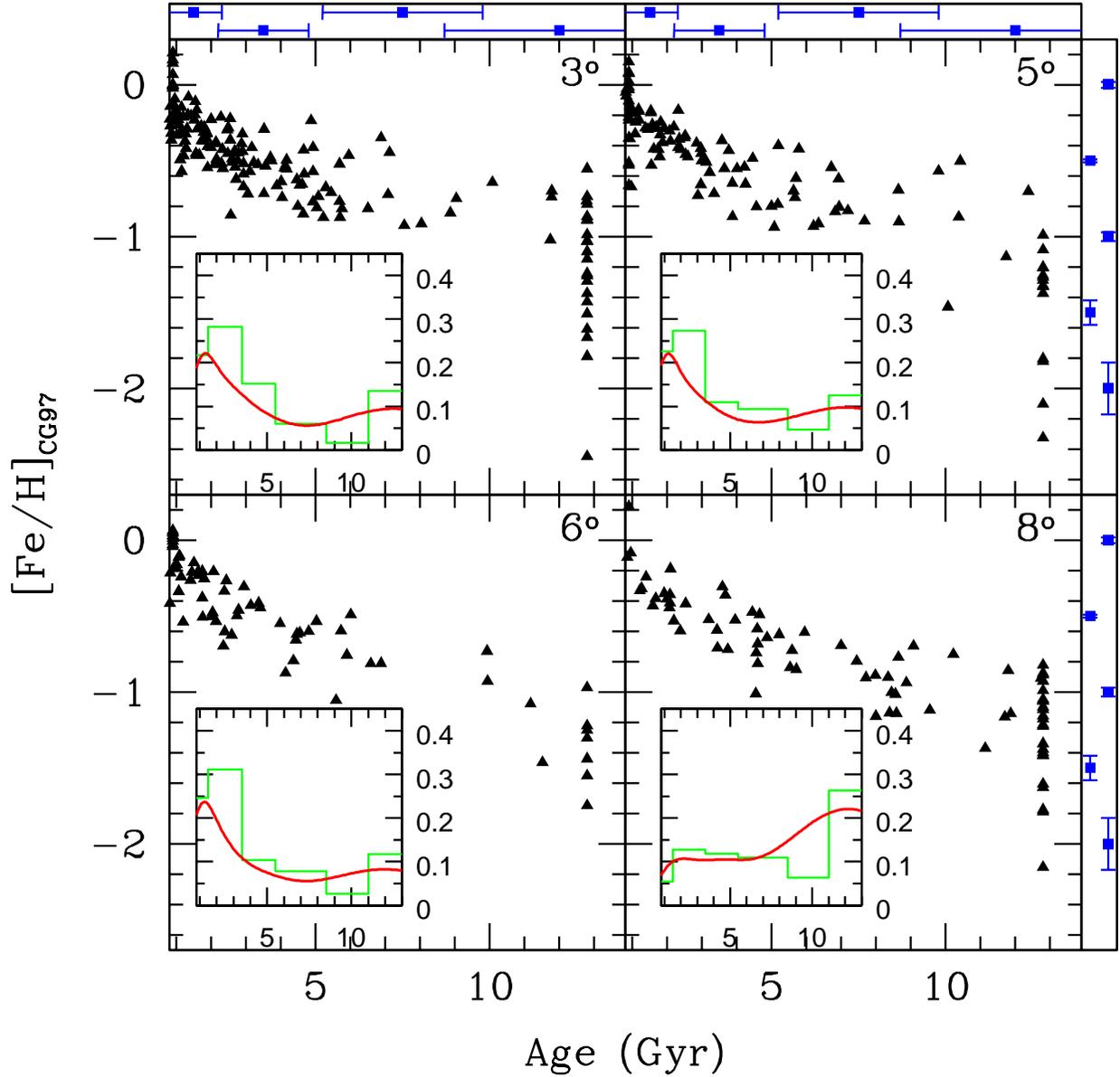}
\caption{Age--metallicity relationships for the four LMC fields in our sample. Inset panels show the age
distribution computed taking into account (\textsl{solid line}) and not (\textsl{histogram}) the age determination uncertainties.
The top panel show the age error in each interval. The left panels show the metallicity
error.
\label{amrfields}}
\end{figure}

The age of each star versus its
metallicity (i.e. the AMR) has been plotted in Figure \ref{amrfields} for each field. The many stars with the same age that appear both at the old (12.9 Gyr) and young (0.8 Gyr) limits are the consequence of the boundary conditions imposed on the ages, and the exact values should not be taken at face value. The age error in each age interval, computed in Section \ref{agedetermination}, is indicated in the top panel. The age distribution for each field has been plotted in inset
panels. The histogram is the age distribution without taking into account 
the age uncertainty. The solid line is the
same age distribution computed by taking into account the age error. To do this, a Gaussian probability
distribution is used to represent the age of each star. The mean and $\sigma$ of the Gaussian are the age obtained for the
star and its error, respectively. The area of each Gaussian is unity. In
stars near the edges, the wings of the distributions may extend further than the age physical limits. In such
cases, we truncated the wing and rescaled the rest of the distribution such that the area remains unity.

\begin{figure}
\epsscale{1}
\plotone{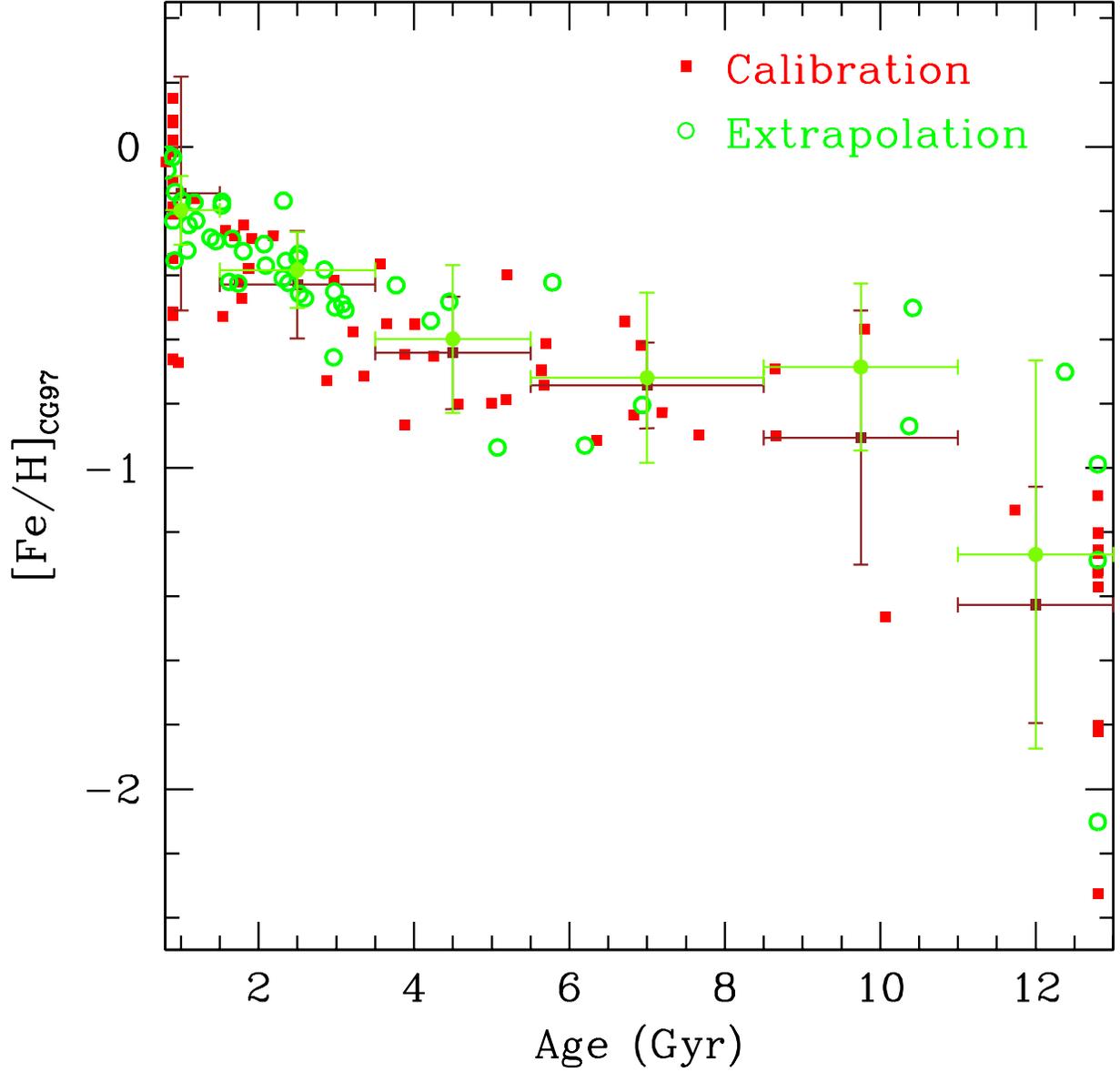}
\caption{AMR of the field situated at 5\arcdeg\ for different star groups. Squares: objects with magnitudes in
the range of the M$_I$--$\Sigma Ca$ calibration obtained in Paper\ II; open circles:
objects in the same field with magnitudes brighter than the cluster calibration stars. The corresponding average metallicity and
dispersion in different age bins are also plotted. \label{testamr}}
\end{figure}

Stars brighter than M$_I$=-3.5 were not used in Paper II to obtain the metallicity calibration of $\Sigma$Ca (Equation \ref{metaleq}). However, we observed stars brighter than M$_I$=-3.5
in the LMC. Therefore, extrapolation of Equation \ref{metaleq} has been used to obtain the
metallicity of these stars, but only for those below the tip of the RGB. It is necessary to check whether this
has introduced any bias in the obtained AMR. Paper\ II demonstrated that
the sequences described by cluster stars used for the calibration in the M$_I$--$\Sigma Ca$ plane are not exactly
linear and  have a quadratic component. In the interval M$_I$=0 -- -3.5 the difference between the
quadratic and linear behaviors are negligible. However, for brighter stars, the deviation from the linear
behavior might be important, resulting in a possible underestimation of the metallicity of these stars. To check this point, in Figure \ref{testamr} we have plotted
as filled squares stars with magnitudes within the range of the calibration.
Open circles are stars whose metallicities have been obtained by extrapolating  
the relationship (-3.5$\geq M_I\geq$-4). The mean metallicity and its dispersion for several age intervals have been also plotted for both cases. The differences between both groups
are smaller than the observed dispersion and therefore we conclude that no strong bias is produced as a
consequence of extrapolating Equation \ref{metaleq} for bright stars. 

Figure \ref{amrfields} shows that the AMR is, within the
uncertainties, very similar for all fields. As expected, the most metal-poor stars in each field are also the oldest 
ones. A rapid chemical enrichment at a very early epoch is followed by a period of very slow metallicity
evolution until around 3 Gyr ago, when the galaxy started another period of chemical enrichment that is still
ongoing. Furthermore, the age histograms for the three innermost fields are similar, although
the total number of stars decreases when we move away from the centre. The outermost field has a lower
fraction of ''young'' (1--4 Gyr) intermediate-age stars. This indicates that its lower mean metallicity is
related to the lower fraction of intermediate-age, more metal-rich stars rather than to a different chemical
enrichment history (for example, a slower metal enrichment).
 
With the aim of obtaining a global AMR for the disk, in the following we will quantitatively address the
question of whether the AMRs of all our disk fields are
statistically the same. We have calculated the mean metallicity of the stars and its dispersion in six age intervals, for each field and for the combination of the four.
The values obtained are listed in Table \ref{testchi2}. To compare the values obtained 
in each field with those for the combined sample, we
performed a $\chi^2$ test such that $\chi^2=\sum_{i=1}^6\frac{(Z_i^{field}-Z_i^{comb})^2}{\sigma_i^2}$,
where $\sigma^2_i$ is the sum of the uncertainties squared in the age bin $i$ of the field and the combined
AMR. The last column in Table \ref{testchi2} shows the values $\chi^2_\nu=\chi^2/\sqrt{5}$. From these values we may conclude that the AMR of each field is equivalent to that of the combined sample to a probability of more than 99 per cent. 

\begin{figure}
\epsscale{1}
\plotone{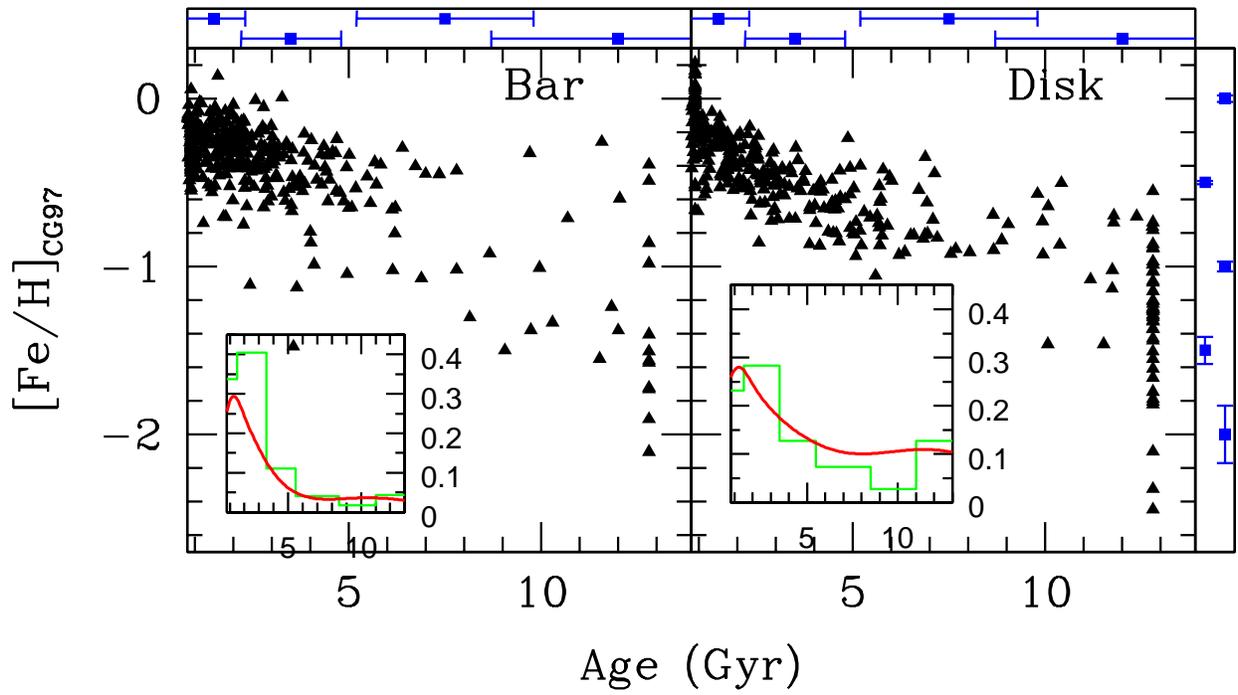}
\caption{Same as Figure \ref{amrfields} but for the bar (\textsl{left}) and the disk fields combined (\textsl{right}).\label{amrbardisk}}
\end{figure}

Now we will compare the global AMR obtained for the disk with that observed in the bar.
The bar stellar ages have been obtained using the relation calculated in Section
\ref{agedetermination}. The age versus metallicity for
each star in the bar (\textsl{left}) and disk (\textsl{right}) have been plotted in Figure
\ref{amrbardisk}. Notice that with a different method and different stellar evolution models, the AMR obtained
here for the bar is quite similar to that obtained by \citet{c05}. We performed a $\chi^2$ test as before between the galaxy disk and bar AMRs. They are the same to within a probability of 90 per cent. However, a visual comparison would indicate that, while in the disk the metallicity has
  risen monotonically over the last few Gyr, this is not the case for
the bar. However, from the comparison of the mean metallicities and the errorbars for stars in this youngest bin (1--3 Gyr) it appears  that this difference is statistically significant.

\begin{figure}
\epsscale{1}
\plotone{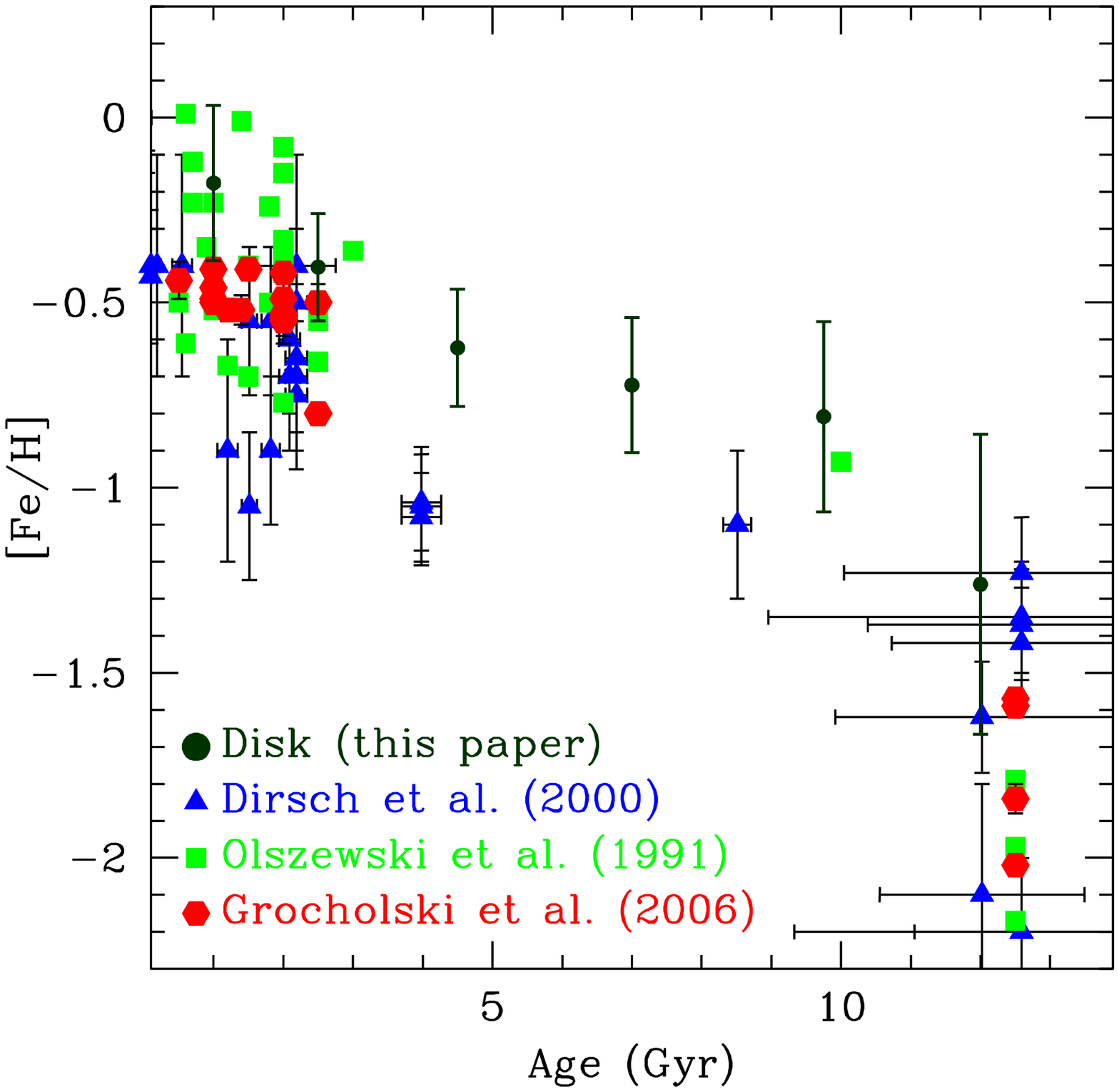}
\caption{LMC clusters AMR by \citet[\textsl{squares}]{ol91}; \citet[\textsl{triangles}]{dirsch00} and \citet[\textsl{hexagons}]{grocholski06}. The mean metallicity in six age bins of our global disk has been plotted (\textsl{filled
points}). Note that the metallicity scales of each work may not be exactly the same.\label{cumulos}}
\end{figure}

The global disk AMR is qualitatively similar to that of the LMC cluster system \citep[e.g.][]{ol91,dirsch00},
except for the lack of intermediate-age clusters (Figure \ref{cumulos}). Clusters show a rapid
enrichment phase around 10 Gyr ago which is also observed in the disk and 
the bar. The second period of chemical enrichment in the last 3 Gyr is also observed both in the field and in the clusters. The field intermediate-age stars may have contributed to
the chemical enrichment at recent times. Note that the \citet{grocholski06} sample has a maximum 
metallicity [Fe/H]=-0.6, i.e. it does not seem to participate on the high 
metallicity tail at young ages that we and the other authors analysing clusters samples find. In 
the three works, there are a number of young clusters with low 
metallicity for their age, which \citet{bekki07} associate with metal-poor gas stripped from the SMC 
due to tidal interaction between the SMC, LMC and the Galaxy over the 
last 2 Gyr. A few stars also have low metallicity for their age in our 
sample, but they represent a minor contribution to the total.

\subsection{The Star Formation History}\label{sfhcal}

More information can be retrieved from the AMRs. In particular
we can also derive an approximate SFH
of the LMC if we can establish that the observed stars are representative of the total population. The number of stars in each age bin can be transformed into
 stellar total mass, after accounting for the
 number of stars with a given age that have
already died. To evaluate this correction 
we have computed a synthetic CMD using IAC-star 
\citep{aparicio04}. We have
assumed a constant SFR and used the relations derived for the chemical enrichment. As we have only observed a fraction of the total stars in the
region of the RGB, we have rescaled the result to the total
number of stars in this region.

\begin{figure}
\epsscale{1}
\plotone{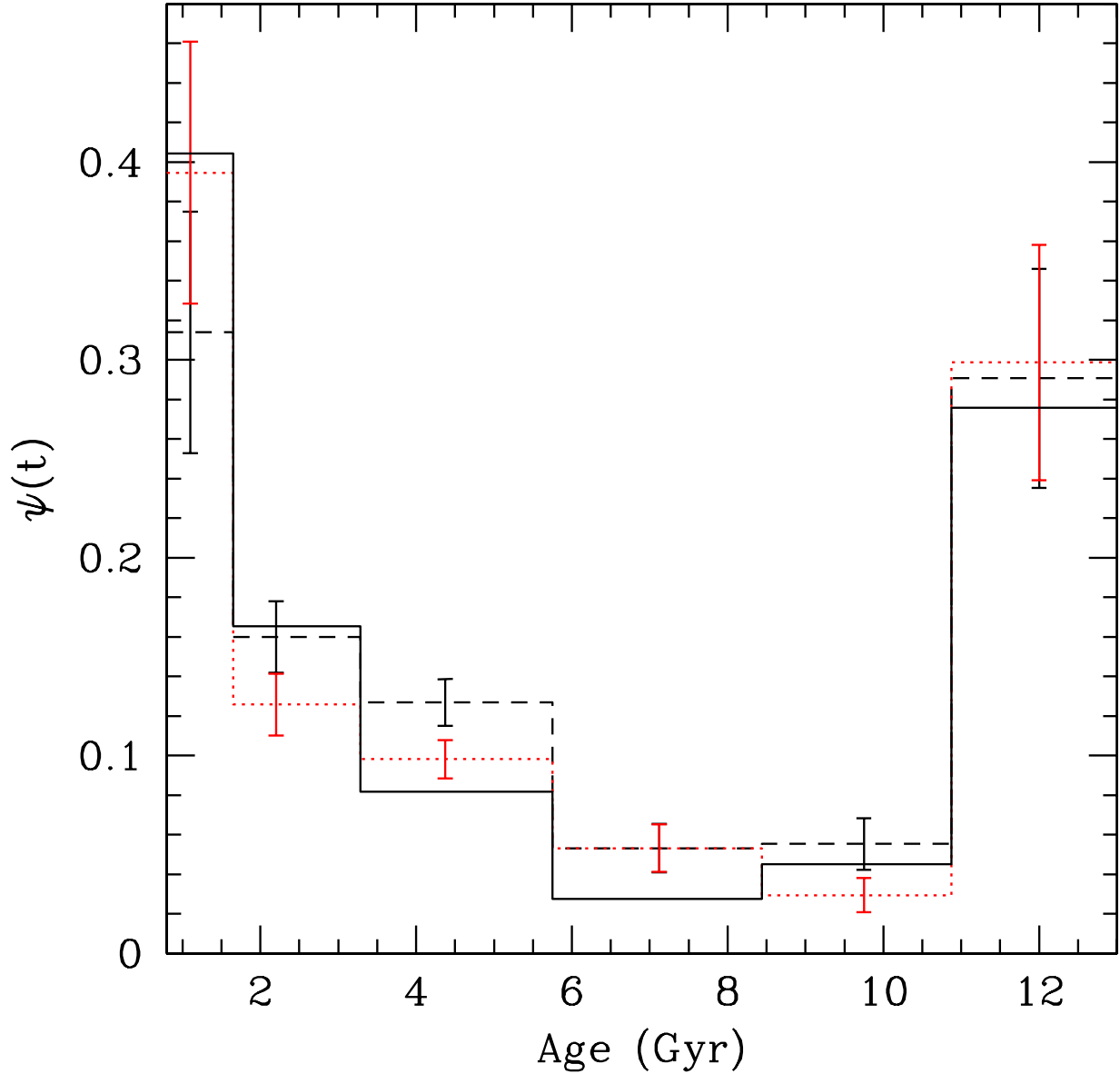}
\caption{Results of the test on the reliability of the SFH computed using our spectroscopic RGB sample. Solid
line: injected SFH; dashed line: mean and sigma of the recovered SFH after the Hydra filter assignment
simulation; dotted line: the same as dashed one but with the age obtained from Equation \ref{rela}. See text for detail.\label{pruebaconf}}
\end{figure}

In order to check whether the procedure used to select the stars observed
spectroscopically introduces any bias in the computed SFH, we have calculated a
synthetic CMD with a known chemical enrichment history and SFR. It was computed such that the number of stars
in the selection region was the same as in the field at 3\arcdeg. The input SFR as a function of age,
$\psi(t)$, is plotted as a
solid line in Figure \ref{pruebaconf}. To check that the
selection criteria did not introduce biases, we have selected objects from this synthetic CMD in the same way
as from the LMC fields (see Figure \ref{dcmbox}). We assigned random positions to each synthetic star and, using the HYDRA
configuration software, we computed 20 test configurations. The number of objects selected in each
configuration, about 115, is similar to the observed stars in the LMC fields. For each test configuration, we
calculated $\psi(t)$ following the procedure described above. The mean of the 20 total tests is the dashed line in
Figure \ref{pruebaconf}. The error bars are the standard deviation of the 20 tests. We repeated the same test, but obtaining the ages of each synthetic star using Equation \ref{rela}. The result is the dotted line in Figure \ref{pruebaconf}.
As we can see in this Figure, the recovered $\psi(t)$
are consistent with the real one. In half of the bins the differences are within the error bars and in the rest
they are within 3$\sigma$. 

\begin{figure}
\epsscale{1}
\plotone{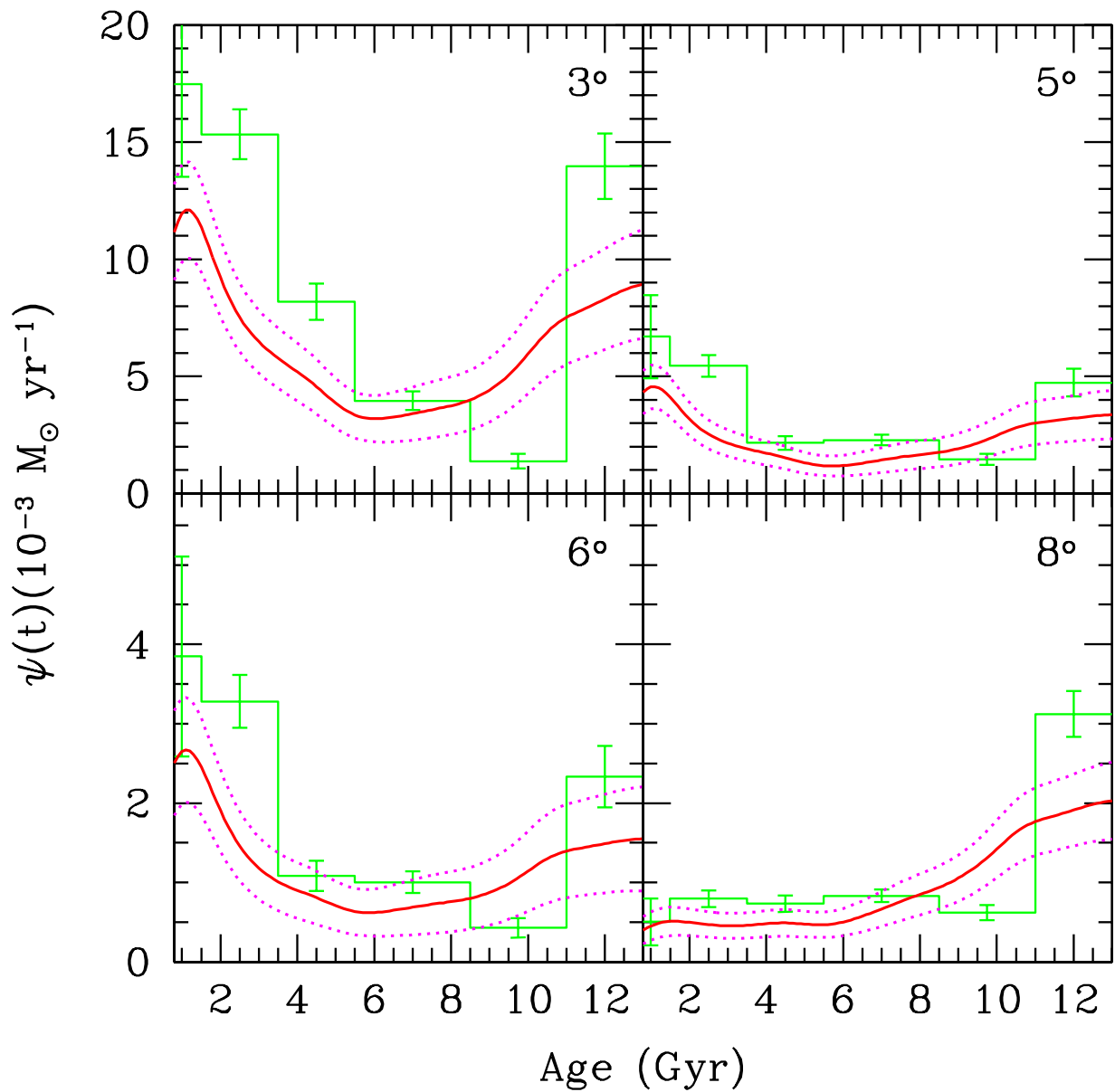}
\caption{SFR, as a function of time, for our four fields. Note that the $y$-axis scale is different in the top and
bottom panels. Histogram is the $\psi(t)$ computed from the age distribution without taking into account the age uncertainty. Solid line is the same but computed by taken into account the age error, while dotted lines are its uncertainty.\label{hfecampos}}
\end{figure}

The $\psi(t)$ derived from our four fields are shown in Figure \ref{hfecampos} (\textsl{solid line}), which has been computed from the age
distribution, taking into account the age error (see Figure \ref{erroredad}). The uncertainty  in the
computed $\psi(t)$ has been estimated as the square root of the SFR value for each age (\textsl{dotted lines}).
However, values of the SFR for different ages are not independent from each other because the integral for the full
age interval is a boundary condition of the solution. In other words, fluctuations above the best solution should be compensated by others below it, and a solution close to, for example, the upper (or the lower) dashed line is very unlike. As a comparison, the $\psi(t)$ computed
without taking into account the age error has also been plotted in 
Figure \ref{hfecampos}. All fields have a first episode of
star formation more than 10 Gyrs ago. Then, $\psi(t)$ decreases until $\sim$5 Gyr ago, when it rises again in the inner regions of the LMC (r$\leq$6\arcdeg). This enhancement of $\psi(t)$ is not observed in the outermost field.

\begin{figure}
\epsscale{1}
\plotone{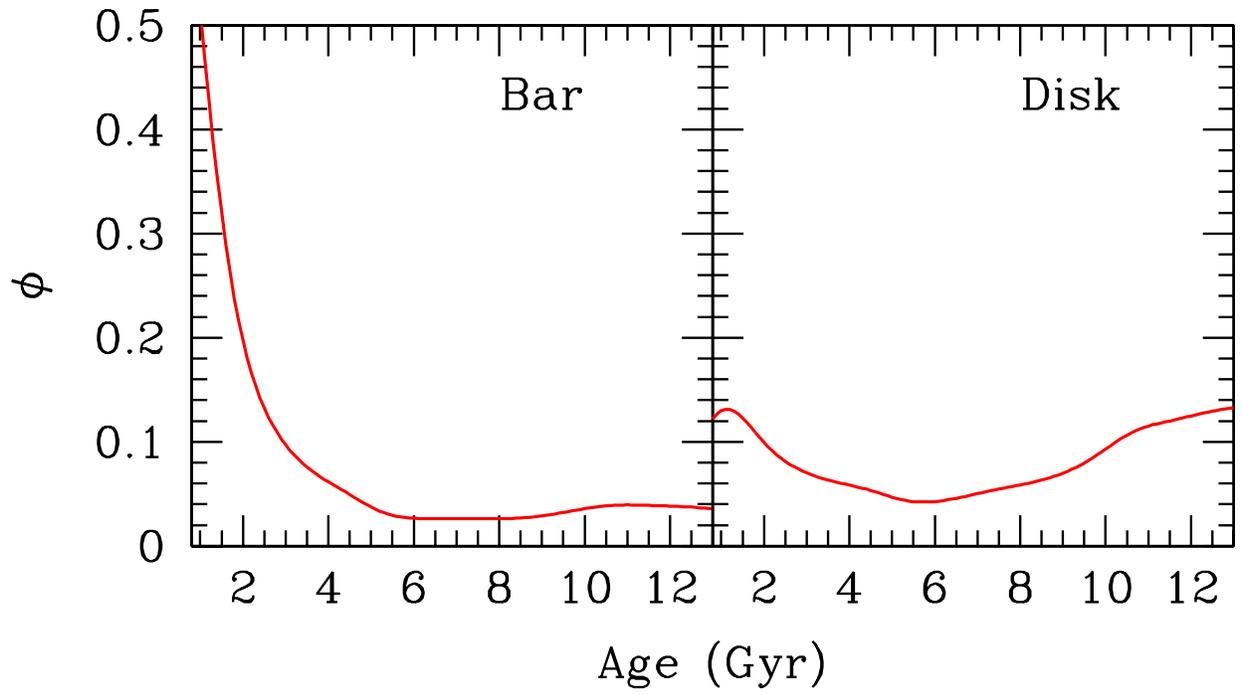}
\caption{SFR as a function of time for the bar (\textsl{left}) and disk (\textsl{right}) derived from the spectroscopic sample.\label{hfe}}
\end{figure}

We followed the same procedure to calculate $\psi(t)$ for the bar stars. The bar $\psi(t)$ is shown in Figure \ref{hfe}. In this case we do not
know the total number of objects in the region where the observed stars were selected. For this reason, we have
plotted the percentage, over the total, that the SFR represents in each age bin. In the right panel of the same
Figure, the $\psi(t)$ of the disk, obtained as a combination of the four fields in our sample, is shown. The bar $\psi(t)$ is 
similar to that derived in small fields from {\itshape HST} deep photometry 
\citep{holtzman99,Smecker-Hane02}, as indeed is expected since the {\itshape HST} fields
are in the same region in which the RGB stars were observed. Both in the bar and in the
disk, the galaxy started forming stars more than 10 Gyr ago, although
in the bar the initial enhancement of $\psi(t)$ is not observed. For intermediate ages, the SFR was low until about 3 Gyr ago, when another increase in the SFR, which is particularly intense in 
the bar, is
observed. This is consistent with previous investigations
\citep[e.g.][]{hardy84,Smecker-Hane02} and with the SFR derived from the clusters. The cluster age gap
coincides with the age interval in which star formation has been less efficient.
  
\subsection{A Chemical Evolution Model for the LMC\label{modelsec}}

In what follows we use the AMR to obtain information on the physical
parameters governing the chemical evolution of the LMC.  As a first
approximation, we will try to reproduce the derived relations with
simple models.

Under the assumption of instantaneous recycling approximation,
following \citet{tinsley80} and \citet{peimbert94}, the
heavy-element mass fraction in the ISM, $Z$, evolves via:

\begin{equation}\label{variaz}
\mu\frac{dZ}{d\mu}=\frac{y(1-R)\psi+(Z_f-Z)f_I}{-(1-R)\psi+(f_I-f_O)(1-\mu)}
\end{equation}

where $f_I$ and $f_O$ are the inflow and outflow rates respectively,
$Z_f$ is the metallicity of the infall gas, $y$ denotes the yield,
$R$ is the mass fraction returned to the interstellar medium by a
generation of stars, relative to the mass locked in stars and stellar
remnants in that generation, $\psi$ is the SFR and $\mu$ is defined
as $\mu=M_g/M_b$, where M$_g$ is the gas mass and M$_b$ is the total
baryonic mass, i.e., the mass participating in the chemical evolution
process. In general, $Z, \mu, \psi, f_I$ and $f_O$ are explicit
functions of time, while $R$ and $y$ may change according to
characteristics of the stellar population like metallicity and initial
mass function, but are usually assumed to be constant for a given
population.

When there are no gas flows in and out of the system ($f_I=f_O=0$),
the model is called a closed-box model and the solution to Equation
\ref{variaz} provides the variation of $Z(t)$ as a function of
$\mu(t)$ via

\begin{equation}
Z(t)=Z_i+y\ln\mu(t)^{-1}
\end{equation}

In an infall scenario, in which gas flows into the system at a rate
proportional to the amount of star formation, $f_I=\alpha(1-R)\psi$,
$\alpha$ being a free parameter, we can integrate Equation
\ref{variaz} for $\alpha\neq1$, assuming that $f_O=0$, and obtaining:

\begin{equation}
Z(t)=Z_i+\frac{y+Z_f\alpha}{\alpha}\left\lbrack 1-
\left(\alpha-\frac{1-\alpha}{\mu(t)}\right)^{-\alpha/(1-\alpha)}\right\rbrack 
\end{equation}

For $\alpha$=1, the solution of Equation \ref{variaz} is

\begin{equation}
Z=Z_i+y\left\lbrack 1-e^{\left(1-\mu^{-1}\right)}\right\rbrack
\end{equation}

In an outflow scenario, in which gas scapes from the system at a rate
$f_O=\lambda(1-R)\psi$, $\lambda$ being a free parameter, we obtain
from Equation \ref{variaz} assuming $f_I=0$, and for ($\lambda\neq1$): 

 \begin{equation}
Z(t)=Z_i+\frac{y}{\lambda-1}\ln\left\lbrack
\frac{\lambda-1}{\mu(t)}-\lambda\right\rbrack 
\end{equation}

Finally, we will consider a model combination of inflow and outflow,
and assume that the system gas flow is proportional to the mass that
has taken part in the star formation: $f=(\alpha-\lambda)(1-R)\psi$,
where we can define $\beta=\alpha-\lambda$. In this case, we can
integrate Equation \ref{variaz} for $\beta\neq$1, to obtain

\begin{equation}
Z(t)=Z_i+\frac{y+Z_f\alpha}{\alpha}\left\lbrack
1-\left(\beta-\frac{1-\beta}{\mu(t)}\right)^{-\beta/(1-\beta)}\right\rbrack 
\end{equation}

In summary, the chemical evolution of the galaxy is given by the yield $y$, the initial metallicity of the gas Z$_i$, the fraction of
gas mass $\mu(t)$, which is a function of $\psi(t)$, and $R$, the mass
fraction returned to the interstellar medium. In the case of
inflow and/or outflow, also by the additional parameters $\alpha$ and
$\lambda$.

We can now apply the former relations to the LMC. In the previous
section we have derived $\psi(t)$ for the LMC bar and disk, the latter
being obtained from the combination of the four fields in our
sample. $\mu$ is computed explicitly as a function of time from $\psi(t)$, the balance of gas flowing to and from the system, and using as a boundary condition the current LMC gas fraction. In the case of the LMC disk, we have assumed the value derived by
\citet{kim98} from observations of H {\footnotesize I}. These authors
derived $M_g=5.2\times10^8 M_{\odot}$ and estimated that the total
mass of the disk is $M_b=2.5\times10^9 M_{\odot}$. With these values,
we find the current gas fraction in the disk, $\mu_f=0.21$. For the
bar we have assumed the value of $\mu=0.08$ given by
\citet{westerlund90}. For the yield, we have assumed the value
obtained by \citet{pagel95} for the solar neighborhood,
$y=0.014$. Assuming the \citet{scalo86} initial mass function,
\citet{maeder93} calculated a returned fraction associated to this
yield of $R=0.44$. Putting all these ingredients together, we will now
explore several scenarios in order to reproduce the AMR of the disk
and the bar.

\begin{figure}
\epsscale{1}
\plotone{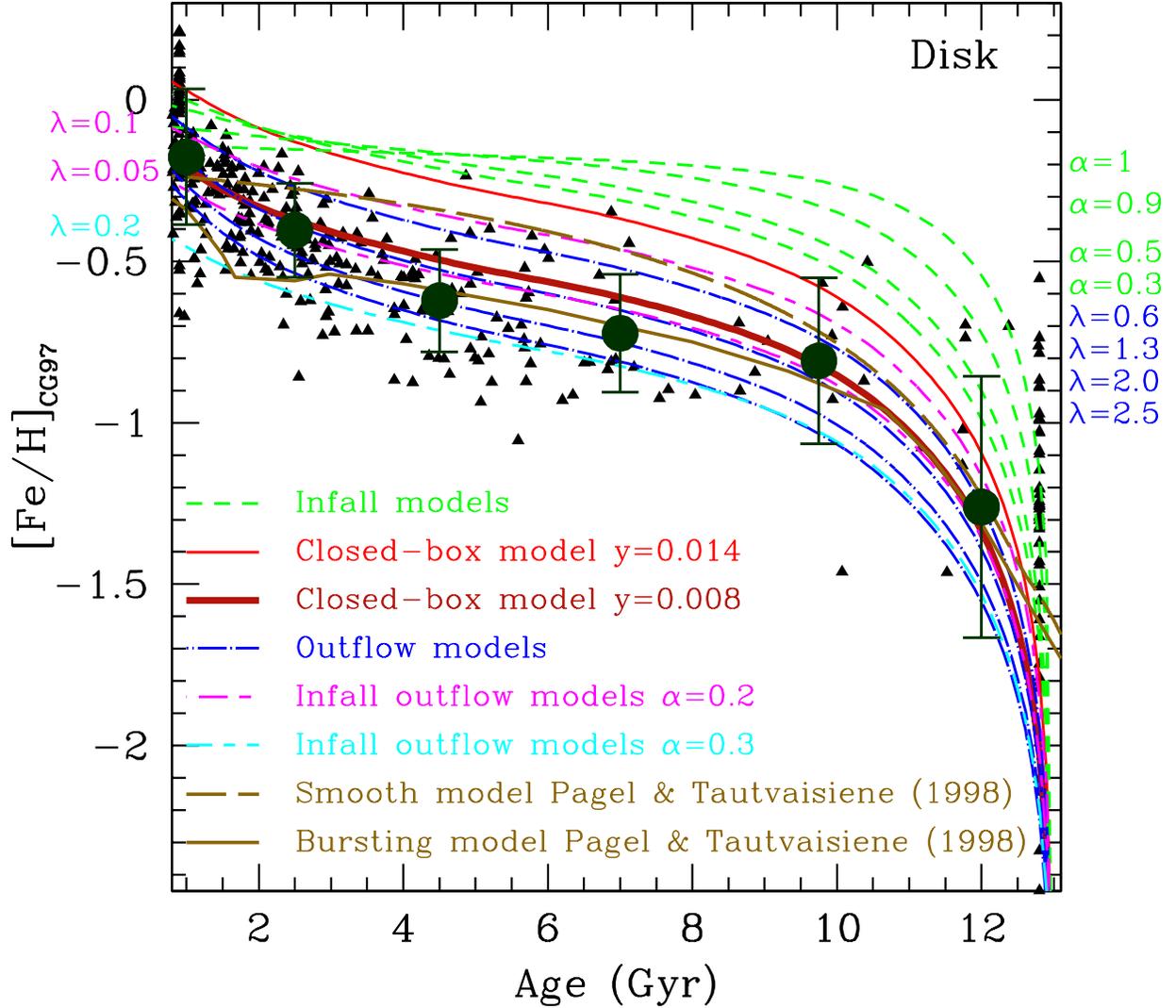}
\caption{Chemical evolution models overplotted on the AMR for the disk. Red solid lines are the closed-box
models with $y=0.008$,  (\textsl{thick line}) and $y=0.014$ (\textsl{thin line}), respectively.
Blue dot-dashed lines are outflow models for $\lambda =$ 0.6, 1.3, 2.0 and 2.5. Green dashed lines are infall models with
$\alpha=$ 1, 0.9, 0.5 and 0.3. Pink long--short dashed lines are models with infall
and outflow for $\alpha=$ 0.2, and $\lambda=$ 0.1 (\textsl{upper}) and 0.05 (\textsl{lower}). Finally, cyan short-long dashed line is a model with inflow
and outflow for $\alpha=$ 0.3 and $\lambda=$ 0.2. All infall and outflow models assume $y=0.014$. The brown solid and dashed lines represent the bursting and smooth models computed by \citet{pagel98}.\label{modeldisk}}
\end{figure}

In Figure \ref{modeldisk} we have plotted the age and metallicity of
each star in the disk, together with the mean metallicity and its
dispersion, for stars in each age bin. In this figure, chemical
evolution models have been superposed for different scenarios and
parameter assumptions (see the figure caption). In the case of the infall models (\textsl{green dashed lines}), we have assumed a zero-metallicity infalling gas, in order to obtain the current
metallicity. They do not reproduce
the general behavior of the observed AMR. The disk AMR is well reproduced by outflow models (\textsl{blue dot-dashed lines}) with a relatively large range of $\lambda$. Finally, the combination of
inflow and outflow (\textsl{pink and cyan long--short dashed lines}) models also reproduce the metallicity
tendency (in this case the model that best matches the observed data
has $\alpha$ = 0.2 and $\lambda$ = 0.05). The previous models have
been computed assuming the yield observed in the solar
neighborhood which is probably too large for the LMC. With a smaller yield, $y=0.008$ (\textsl{red thick solid line}), the observed
AMR can be reproduced by a closed-box model. This yield would correspond to
stars with $Z=0.001$ \citep[see][for details]{maeder93}, which is
slightly low compared with the mean LMC metallicity. Bursting and smooth models computed by \citet{pagel98} have been
plotted as comparison (\textsl{brown lines}). The bursting model was computed assuming two burst of star formation, one
about 12 Gyr ago and a strong one about 3 Gyr ago, and some fraction of infall and outflow. Note that their
predicted values agree with our mean metallicity for each age bin, within the errorbars, specially at old and
intermediate--age. It also qualitatively agrees with the episode of faster chemical evolution with started
around $\sim$2 Gyr ago. The smooth model was computed assuming a constant star formation rate and does not match with the observations.

\begin{figure}
\epsscale{1}
\plotone{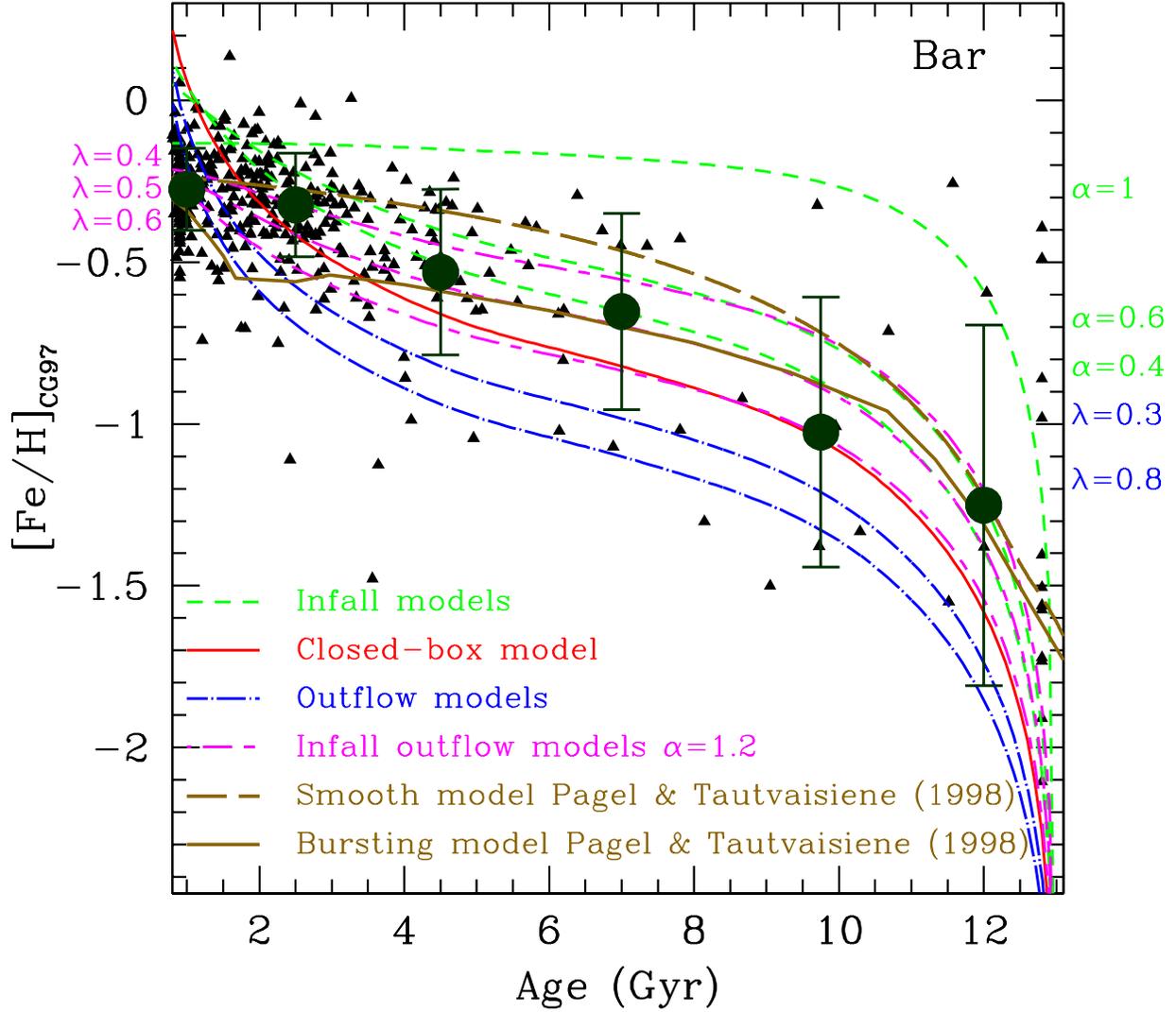}
\caption{As Figure \ref{modeldisk} for the bar. Red solid line is a 
closed-box model. Blue dot-dashed lines are outflow models for
$\lambda=$ 0.8 and 0.3. Green short-dashed lines are infall models with $\alpha=$ 1, 0.6 and
0.4. Pink long--short dashed lines are models with inflow
and outflow for $\alpha=$ 1.2 and $\lambda=$ 0.4, 0.5 and 0.6. All models assume y=0.014. As in the previous
figure, brown lines represent the models computed by \citet{pagel98}.\label{modelbar}}
\end{figure}

In the case of the bar (Figure \ref{modelbar}), the very slowly rising 
metallicity, almost constant over the last few Gyr, is best
reproduced by the combination of models with inflow and outflow with
infall parameter $\alpha$ = 1.2 and $\lambda$ = 0.4-0.6 (\textsl{pink long--short dashed lines}). Infall
models (\textsl{green dashed lines})  with $\alpha$ = 0.4-0.6 marginally reproduce the observed trend, but they predict higher metallicities than observed in the last few Gyr, unless a smaller yield would be
assumed. Finally, outflow and closed-box models predict a rise of metallicity in the last few Gyr more steep than observed. The \citet{pagel98} bursting model agrees worse with the bar AMR than with that of the disk, specially in the youngest age bins.

\section{Conclusions} \label{conclusions}

Using infrared spectra in the CaT region, we have obtained metallicities and radial velocities for a sample of stars in four LMC fields.
Metallicities have been calculated using the relationships between the equivalent width of the CaT lines, $\Sigma Ca$, and metallicity derived in
Paper II. In addition, we have estimated the age of each star using a relationship derived from a synthetic CMD which,
from the color, magnitude and metallicity of a star, allow us to estimate its age. The main results of this paper are:
\begin{itemize}
\item The velocity distribution observed in each field agrees with the rotational thick disk kinematics of the
LMC. The velocity dispersion is slightly larger for the most metal-poor stars. However, the  values obtained do not indicate
the presence of a kinematically hot halo.
\item The metallicity distribution of each field has a well defined peak with a tail toward low
metallicities. The mean metallicity is constant until the field at 6\arcdeg\@ ([Fe/H]$\sim$-0.5 dex), and  is a factor two more
metal-poor for the outermost field ([Fe/H]$\sim$-0.8 dex).
\item The AMR observed in each disk field is compatible with a single global relationship for the disk. We conclude that the outermost field is more metal-poor on average because it contains a lower fraction of relatively young stars (age$\leq$5 Gyr), which are also more metal-rich.
\item The disk AMR shows a prompt initial chemical enrichment. Subsequently, the metallicity increased very
slowly until about 3 Gyr ago, when the rate of metal enrichment increased again. This AMR is
similar to that of the cluster system, except for the lack of clusters with ages between 3 and 10 Gyr. The
recent fast enrichment observed in the disk and  in the cluster system is not observed in the bar.
\item The $\psi(t)$ of the three innermost fields show a first episode of star formation until about 10 Gyr ago, followed by a
period with a low SFR until $\sim$5 Gyr ago, when the SFR increases, and reaches its highest values $\sim$2-3 Gyr ago. The outermost field does not show the recent increase of SFR. The second main episode is also observed, and is more prominent in the bar, where an increased SFR at old ages is not observed. The lower SFR between 5 and 10 Gyr ago is probably related
to the age-gap observed in the clusters.
\item Under the assumption of a solar yield, the disk AMR is well reproduced either by a chemical evolution model with outflow with
$\lambda$ between 1 and 2, so the disk losses the same amount of gas that has taken part in the star formation, or by models combining infall and outflow with $\alpha$=0.2 and
$\lambda$=0.05, which means that the galaxy is almost a closed-box system. With
a smaller yield, the AMR could also be reproduced with a closed-box
model. The bar AMR is well reproduced by models with a combination of inflow and outflow with
$\alpha$=1.2 and $\lambda$ between 0.4 and 0.5. This suggests that the amount of infalling gas was larger than the amount that participated in the star formation in the bar, and also that the amount of gas that escaped the bar was 50\% of the total that participated in star formation.
\end{itemize}

\section{Discussion} \label{discussion}
The main result in this paper is that all our fields, covering
galactocentric radius from 3\arcdeg\@ to 8\arcdeg\@ from the center (2.7 to 7.2 kpc) share a very similar AMR. The mean metallicity is very
similar in the three innermost fields ([Fe/H] $\simeq$-0.5 dex), and it
is a factor of $\simeq 2$ smaller in the outermost field ([Fe/H] $\simeq$-0.8 dex). Because the
AMRs are the same, this has to be related with the lower fraction of
relatively young stars (age $\le$ 5 Gyr), which are also more metal-rich, in the outer part. We find, therefore, a change in the age
composition of the disk population beyond a certain radius ($\simeq$ 6
kpc; note however that in this study we are only sensitive to
populations older than $\simeq$ 0.8 Gyr, and that differences among
fields for younger ages could also exist), while the chemical
enrichment history seems to be shared by all fields.

The comparison of the AMR observed in the disk with simple chemical evolution models suggests 
that the LMC is most likely losing some of its gas. This is in agreement with the work by 
\citet{nidever07} which suggests that the main contribution to the Magellanic Stream comes from the 
LMC instead of from the SMC as it was believed until now \citep[e.g.][]{putman03}. Alternatively, the LMC 
could be an almost closed-box system with small gas exchanges ($\alpha$=0.2, $\lambda$=0.05). This could be in
agreement with the models by \citet{bekki07} which suggest that the LMC has received metal-poor gas 
from the SMC in the last 2 Gyr. In our field we observed some stars with a low metallicity to which we assign an age younger than 2 Gyr. However, the uncertainty in the age determination is large.

We have also obtained the AMR of the LMC bar from the data by \citet{c05}. The
bar AMR differs from the one of the disk in the last 5 Gyr: while in
the disk the metallicity has increased in this time, in the bar it has
remained approximately constant. This feature is best reproduced by
models combination of outflow and a relatively large infall of pristine gas.  Models with a smaller yield,
but with an infall of previously enriched gas also reproduce the
observed bar AMR. This would be in agreement with the prediction that a typical bar instability pushes the gas of the disk towards the center \citep{sellwood93}, so the infall gas is expected to be previously pre-enrichment by the disk chemical evolution. This would be also the case in the scenario suggested by \citet{bekki05}, in which the bar would have formed from disk material as a consequence of tidal interactions between the LMC, the SMC and the Milky Way about 5 Gyr ago, the moment when the bar AMR differs from that of the disk.  Also, a bar is expected to destroy any metallicity gradient within a certain radius, as it is observed. 

\citet{gshpz04} derived the LMC surface brightness profile using
deep resolved star photometry of the four fields in the current
spectroscopic study, and found that it remains exponential to a radius
of 8\arcdeg\@ ($\simeq$ 7 kpc), with no evidence of disk
truncation. Combining this information with that on the deep CMD of
the outermost field, which contains a large fraction of
intermediate-age stars, they concluded that the LMC disk extends (and
dominates over a possible halo) at a distance of at least 7 kpc from
its center.  In the present study, the kinematics of the stars in
these four fields confirm this conclusion: the velocity dispersion of
all four fields is similar, around 20 km/sec. If the stars are binned
by metallicity, the velocity dispersion of the most metal-poor bin is
slightly larger than that of the metal-richer bins ($\sigma_V
\simeq 25$ km/sec as opposed to $\simeq 20$ km/sec) but still not
large enough to indicate the presence of a halo, even one formed as a
consequence of the interaction with the Galaxy. In this case, \cite{bekki04} predicted a velocity dispersion $\simeq$ 40 km/s at a
distance of 7.5 kpc from the LMC center (similar to the distance of
our outermost field). Other authors have also failed to find a
kinematically hot halo \citep{freeman83, schommer92, graff00,
zhao03}. The first evidence of the presence of a kinetically hot, old
spheroidal population was reported by \citet{hughes91} using a sample
of long period variables, which are related to an old stellar
population. Recently, \citet{miniti03} observed spectroscopically a
sample of 43 bar RR Lyrae stars, and obtained a large velocity dispersion
(53 $\pm$ 10 km s$^{-1}$) for them. It is possible that our failure
(and that of other authors) to find evidence of a hot stellar halo is
related with a low contrast of the halo population with respect to the
disk one, even at large galactocentric radius as our outermost field.

\acknowledgments

AA, CG, and RC acknowledge the support from the Spanish 
Ministry of Science and Technology (Plan Nacional de Investigaci\'on Cient\'{\i}fica,
Desarrollo, e Investigaci\'on Tecnol\'ogica, AYA2004-06343) and from 
the Instituto de Astrof\'{\i}sica de Canarias (grants P3/94 and
3I1902). RZ acknowledges the support of National Science Foundation grant AST05-07362. This work has made use
of the IAC-STAR Synthetic CMD computation code. IAC-STAR is supported and maintained by the computer division
of the Instituto de Astrof\'{\i}sica de Canarias.


%% copy editing.  Individual instruments can be provided in parentheses,
%% after the keyword, but they will not be verified.

Facilities: \facility{CTIO(HYDRA)}.

\clearpage

\clearpage

\clearpage

\clearpage

\clearpage

\clearpage

\clearpage

\clearpage

\clearpage

\clearpage

\clearpage

\clearpage

\clearpage

\clearpage

%\clearpage

%\clearpage

%\clearpage

\clearpage

\clearpage

\clearpage

\begin{deluxetable}{ccccccccc}
\tabletypesize{\scriptsize}
%%\rotate
\tablecaption{LMC RGB stars observed\label{starobs}}
\tablewidth{0pt}
\tablehead{
\colhead{$\alpha_{2000}$} & \colhead{$\delta_{2000}$} & \colhead{$\Sigma$ Ca} & \colhead{$\sigma_{\Sigma Ca}$} & \colhead{V} & \colhead{I} & \colhead{$V_r (km s^{-1}$)} & \colhead{$\sigma_{V_r} (km s^{-1}$)} & \colhead{Comments} 
}
\startdata
05:08:53.92 & -66:49:55.1 & 7.8 & 0.8 & 17.67 & 16.36 & 287.3 & 0.7 &              \\
05:08:55.21 & -66:47:34.7 & 8.7 & 0.2 & 16.66 & 14.70 & 240.9 & 0.6 &              \\
05:08:53.89 & -67:02:50.0 & 2.4 & 0.2 & 17.19 & 14.25 &  23.0 & 0.6 & No member   \\
05:08:55.40 & -66:53:45.4 & 8.9 & 0.4 & 16.79 & 15.09 & 292.3 & 0.5 &              \\
05:08:58.58 & -66:45:49.2 & 9.4 & 0.2 & 16.00 & 13.99 & 291.9 & 0.5 &  \\
\enddata
\tablecomments{Table \ref{starobs} is published in its entirety in the electronic edition of The Astronomical 
Journal. A portion is shown here for
guidance regarding its form and content}

\end{deluxetable}

%\clearpage
\begin{deluxetable}{ccc}
\tabletypesize{\scriptsize}
%%\rotate
\tablecaption{Radial velocities and velocity dispersion for each field.\label{radialvelocity}}
\tablewidth{0pt}
\tablehead{
\colhead{Field} & \colhead{$\langle V_r\rangle (km s^{-1})$} & \colhead{$\sigma_{V} (km s^{-1})$} 
}
\startdata
Bar & 260 & 24 \\
3\arcdeg & 269 & 27 \\
5\arcdeg & 278 & 20 \\
6\arcdeg & 293 & 15 \\
8\arcdeg & 282 & 14 \\
\enddata
\end{deluxetable}

\begin{deluxetable}{cc}
\tabletypesize{\scriptsize}
\tablecaption{Line and continuum bandpasses
\label{bandastable}}
\tablewidth{0pt}
\tablehead{
\colhead{Line Bandpasses (\AA)} & \colhead{Continuum bandpasses (\AA)}}
\startdata
8484-8513 & 8474-8484\\
8522-8562 & 8563-8577\\
8642-8682 & 8619-8642\\
\nodata & 8799-8725\\
\nodata & 8776-8792\\
\enddata
\end{deluxetable}

\begin{deluxetable}{lcccccccc}
\tabletypesize{\scriptsize}
%%\rotate
\tablecaption{Coefficients of Equation \ref{rela}.\label{tableage}}
\tablewidth{0pt}
\tablehead{
\colhead{Model} & \colhead{a} & \colhead{b} & \colhead{c} & \colhead{d} & \colhead{f}  & \colhead{g} & \colhead{h} & \colhead{$\sigma$}
}
\startdata
BaSTI & 2.57$\pm$0.10 & 9.72$\pm$0.15 & 0.70$\pm$0.003 & -1.51$\pm$0.007 & -3.86$\pm$0.08 &-0.19$\pm$0.007 &
0.49$\pm$0.01 & 0.38 \\
Padova & 1.69$\pm$0.07 & 10.6$\pm$0.1 & 0.75$\pm$0.002 & -1.87$\pm$0.007 & -4.14$\pm$0.06 &-0.24$\pm$0.006 & 0.51$\pm$0.009 &0.37 \\
\enddata
\end{deluxetable}

\clearpage

\begin{deluxetable}{ccccc}
\tabletypesize{\scriptsize}
%%\rotate
\tablecaption{Velocity dispersion for each metallicity bin.\label{vrmetaltable}}
\tablewidth{0pt}
\tablehead{
\colhead{[Fe/H]} & \colhead{Age (Gyr)} & \colhead{N$_*$} & \colhead{$\langle V_r\rangle (km s^{-1})$} & \colhead{$\sigma_{V} (km s^{-1})$}
}
\startdata
$\geq$-0.5 & $<$3 & 226 & 0.1 & 20.5\\
-0.5 to -1 & 2-10 & 178 & 1.9 & 20.5\\
$\leq$-1 & $>$10 & 100 & 1.8 & 26.4\\
\enddata
\end{deluxetable}

%\clearpage

\begin{deluxetable}{ccc}
\tabletypesize{\scriptsize}
%%\rotate
\tablecaption{Mean values of metallicity distributions.\label{metallicitybin}}
\tablewidth{0pt}
\tablehead{
\colhead{Field} & \colhead{$\langle$[Fe/H]$\rangle$} & \colhead{$\sigma_{[Fe/H]}$} 
}
\startdata
Bar & -0.39 & 0.19 \\
3\arcdeg & -0.47 & 0.31 \\
5\arcdeg & -0.50 & 0.37 \\
6\arcdeg & -0.45 & 0.31 \\
8\arcdeg & -0.79 & 0.44 \\
\enddata
\end{deluxetable}

\begin{deluxetable}{cccccccc}
\tabletypesize{\tiny}
\tablecaption{Average metallicity in six age bins.\label{testchi2}}
\tablewidth{0pt}
\tablehead{
\colhead{Field} & \colhead{$\langle[Fe/H]_{\leq1.5}\rangle$} &  \colhead{$\langle[Fe/H]_{1.5-3.5}\rangle$}
& \colhead{$\langle[Fe/H]_{3.5-5.5}\rangle$} & \colhead{$\langle[Fe/H]_{5.5-8.5}\rangle$} & \colhead{$\langle[Fe/H]_{8.5-11}\rangle$} & 
\colhead{$\langle[Fe/H]_{\geq11}\rangle$} & \colhead{$\chi^2_\nu$} \\ 
}
\startdata
Bar & -0.27$\pm$0.13 & -0.32$\pm$0.16 & -0.53$\pm$0.26 & -0.65$\pm$0.30 & -1.02$\pm$0.41 & -1.25$\pm$0.56 &
... \\
Disk\tablenotemark{a} & -0.17$\pm$0.21 & -0.39$\pm$0.15 & -0.60$\pm$0.16 & -0.71$\pm$0.18 & -0.80$\pm$0.27 & -1.25$\pm$0.41 &
... \\
3\arcdeg & -0.16$\pm$0.17 & -0.40$\pm$0.14 & -0.58$\pm$0.15 & -0.67$\pm$0.20 & -0.69$\pm$0.07 & -1.13$\pm$0.43 & 0.05\\
5\arcdeg & -0.16$\pm$0.28 & -0.39$\pm$0.15 & -0.63$\pm$0.18 & -0.72$\pm$0.16 & -0.83$\pm$0.35 & -1.39$\pm$0.42 & 0.02\\
6\arcdeg & -0.15$\pm$0.16 & -0.38$\pm$0.16 & -0.60$\pm$0.16 & -0.75$\pm$0.20 & -0.83$\pm$0.14 & -1.34$\pm$0.24 & 0.02\\
8\arcdeg & -0.16$\pm$0.21 & -0.45$\pm$0.13 & -0.62$\pm$0.20 & -0.89$\pm$0.16 & -0.92$\pm$0.18 & -1.25$\pm$0.32 & 0.16\\
\enddata
\tablenotetext{a}{Combination of the four fields.}
\end{deluxetable}

\end{document}